\documentclass[12pt,a4paper]{article}

\usepackage{epsfig}
\usepackage{a4wide}
\usepackage{amsmath}
\usepackage{bm}
\usepackage{amssymb}
\usepackage{graphicx}
\usepackage{slashed}

\newcommand{\agt}{\rlap{\lower 3.5 pt \hbox{$\mathchar \sim$}} \raise 1pt
 \hbox {$>$}}
\newcommand{\alt}{\rlap{\lower 3.5 pt \hbox{$\mathchar \sim$}} \raise 1pt
 \hbox {$<$}}

\newcommand{\HS}{S}

\newcommand{\BS}{\mathbb S}
\newcommand{\HP}{\mathrm{H}}
\newcommand{\HPL}{\mathrm{H}}

\newcommand{\w}{\omega}
\newcommand{\pqqp}{p_{qq}(x)}
\newcommand{\pqqm}{p_{qq}(-x)}
\newcommand{\mm}{\varepsilon}

\newcommand{\ca}{C_A}
\newcommand{\cf}{C_F}

\newcommand{\nf}{n_f}

\newcommand{\NP}{\raisebox{2.8pt}{${\footnotesize{\ensuremath\pmb{\diagup}}}$}\!\!\,\!\!\!\!\mathcal{P}}

\def\D#1{D_#1}

\def\z#1{\zeta_#1}

\newcommand{\sign}{\mathop{\mathrm{sign}}\nolimits}

\newcommand{\nr}{N_R}

\def\frct#1#2{\mbox{\small{$\displaystyle\frac{#1}{#2}$}}}

\catcode`@=11
\newcount\@tempcntc
\def\@citex[#1]#2{\if@filesw\immediate\write\@auxout{\string\citation{#2}}\fi
  \@tempcnta\z@\@tempcntb\m@ne\def\@citea{}\@cite{\@for\@citeb:=#2\do
    {\@ifundefined
       {b@\@citeb}{\@citeo\@tempcntb\m@ne\@citea\def\@citea{,}{\bf
?}\@warning
       {Citation `\@citeb' on page \thepage \space undefined}}%
    {\setbox\z@\hbox{\global\@tempcntc0\csname b@\@citeb\endcsname\relax}%
     \ifnum\@tempcntc=\z@ \@citeo\@tempcntb\m@ne
       \@citea\def\@citea{,}\hbox{\csname b@\@citeb\endcsname}%
     \else
      \advance\@tempcntb\@ne
      \ifnum\@tempcntb=\@tempcntc
      \else\advance\@tempcntb\m@ne\@citeo
      \@tempcnta\@tempcntc\@tempcntb\@tempcntc\fi\fi}}\@citeo}{#1}}
\def\@citeo{\ifnum\@tempcnta>\@tempcntb\else\@citea\def\@citea{,}%
  \ifnum\@tempcnta=\@tempcntb\the\@tempcnta\else
   {\advance\@tempcnta\@ne\ifnum\@tempcnta=\@tempcntb \else
\def\@citea{--}\fi
    \advance\@tempcnta\m@ne\the\@tempcnta\@citea\the\@tempcntb}\fi\fi}
\catcode`@=12

\begin{document}

\boldmath
\title{Four-loop anomalous dimension of flavor non-singlet quark operator of twist two and Lorentz spin $N$ for general gauge group: transcendental part}
\unboldmath

\author{B.~A. Kniehl, V.~N.~Velizhanin\\
  {\normalsize II. Institut f\"ur Theoretische Physik, Universit\"at Hamburg,}\\
  {\normalsize Luruper Chaussee 149, 22761 Hamburg, Germany}
}
\date{}

\maketitle

\begin{abstract}
Both for quark flavor asymmetry and valence, we consider the anomalous dimension of the non-singlet twist-two quark operator of arbitrary Lorentz spin $N$ at four loops in SU($n_c$) color gauge theory and present its term proportional to $\zeta(3)$ in closed form.
These results have been extracted from published Mellin moments, for $N=1,\ldots,16$ and $N=3,\ldots,15$, respectively, by analytic reconstruction using advanced methods of number theory.
Via Mellin transformation, we obtain the exact functional forms in $x$ of the respective pieces of Dokshitzer--Gribov--Lipatov--Altarelli--Parisi splitting functions.
This allows us to reduce the theoretical uncertainties in the approximations of these splitting functions otherwise amenable from the first few low-$N$ values.
\medskip

\end{abstract}

\newpage

\section{Introduction}

Parton distribution functions (PDFs) are indispensable ingredients for calculations of cross sections for high-energy collisions with hadrons \cite{Bjorken:1969ja,Collins:1989gx}.
Present and future high-precision measurements at the CERN Large Hadron Collider (LHC) and the BNL Electron-Ion Collider (EIC) rapidly boost the benchmarks for uncertainties in PDFs.
In turn, our knowledge of the Dokshitzer--Gribov--Lipatov--Altarelli--Parisi (DGLAP) equations \cite{Gribov:1972ri,Gribov:1972rt,Altarelli:1977zs,Dokshitzer:1977sg}, which drive the factorization scale evolution of PDFs, must be deepened accordingly.
This implies that the splitting functions $P_{ij}(x)$, which make up the DGLAP evolution kernels, must be pushed to higher orders of perturbation theory,
\begin{equation}
P_{ij}(x)=\sum_{n=0}^{\infty}a_s^{n+1}P_{ij}^{(n)}(x)\,,
\end{equation}
where $a_s=\alpha_s(\mu)/(4\pi)$ with $\alpha_s(\mu)$ being the strong-coupling constant.

Given the PDFs of $f$-flavored quarks and antiquarks and the gluon, $q_f$, $\bar{q}_f$, and $g$, it is advantageous to organize the quark sector in terms of flavor asymmetries $q_{\mathrm{ns},ff^\prime}^{\pm}=(q_f\pm \bar{q}_f)-(q_{f^\prime}\pm \bar{q}_{f^\prime})$ and valence $q_{\mathrm{ns}}^{\mathrm{v}}=\sum_f(q_f-\bar{q}_f)$ in non-singlet configuration, and flavor singlet $q_{\mathrm{s}}=\sum_f(q_f+\bar{q}_f)$.
Then the DGLAP evolution proceeds separately for $q_{\mathrm{ns},ff^\prime}^{\pm}$ and $q_{\mathrm{ns}}^{\mathrm{v}}$, leaving a $2\times2$ matrix equation for $q_{\mathrm{s}}$ and $g$.
This involves seven distinct splitting functions, $P_{\mathrm{ns}}^{\pm}$, $P_{\mathrm{ns}}^{\mathrm{v}}$, $P_{qq}$, $P_{qg}$, $P_{gq}$, $P_{gg}$.

In principle, splitting functions may be directly calculated from the Feynman diagrams of the $j\to i$ parton transitions including their radiative corrections.
An alternative, actually more powerful, approach relies on the fact that they are related via Mellin transformation,
\begin{equation}
\gamma_{ij}(N)=-\int_0^1 dx\ x^{N-1} P_{ij}(x)\,,
\label{eq:mel}
\end{equation}
to the anomalous dimensions
\begin{equation}
\gamma_{ij}(N)=\sum_{n=0}^{\infty}a_s^{n+1}\gamma_{ij}^{(n)}(N)\,,
\end{equation}
of appropriately defined local composite operators of twist two and Lorentz spin $N$, constructed from quark fields $\psi$, gluon field strength tensors $G^{\mu\nu}$, and a definite number of covariant derivatives $D^\mu$, namely $N-1$ for quark operators and $N-2$ for gluon operators \cite{Dokshitzer:1977sg}.
The block diagonal structure of the splitting functions carries over to Mellin space via Eq.~\eqref{eq:mel}.
Specifically, the flavor non-singlet and singlet quark operators and the gluon operators are defined as
\begin{eqnarray}
\mathcal{O}^{a,\{\mu_1\ldots\mu_N\}}&=&\bar\psi\lambda^a
\gamma^{\{\mu_1}{\mathcal D}^{\mu_2}\ldots {\mathcal D}^{\mu_N\}}\psi\,,
\label{eq:nonsin}\\
\mathcal{O}^{\psi,\{\mu_1\ldots\mu_N\}}&=&\bar\psi
\gamma^{\{\mu_1}{\mathcal D}^{\mu_2}\ldots {\mathcal D}^{\mu_N\}}\psi\,,
\\
\mathcal{O}^{G,\{\mu_1\ldots\mu_N\}}&=&G_\mu^{\phantom{\mu}\{\mu_1}
{\mathcal D}^{\mu_2}\ldots {\mathcal D}^{\mu_{N-1}}G^{\mu_N\}\mu}\,,
\end{eqnarray}
where the brackets imply total symmetrization in the Lorentz indices and removal of all traces, and $\lambda^a$ with $a=3,8,\ldots,n_f^2-1$ are the diagonal generators of the flavor group SU$(n_f)$ with $n_f$ being the number of quasi-massless quark flavors.
Having gathered a sufficiently large set of low-$N$ moments of a given anomalous dimension $\gamma_{ij}^{(n)}(N)$, it is possible to analytically reconstruct the all-$N$ result, which is equivalent to establishing the exact $x$ dependence of the respective splitting function $P_{ij}^{(n)}(x)$.

The leading-order terms, $P_{ij}^{(0)}(x)$ and $\gamma_{ij}^{(0)}(N)$, were derived diagrammatically in the pioneering works of Ref.~\cite{Gribov:1972ri,Gribov:1972rt,Altarelli:1977zs,Dokshitzer:1977sg,Gross:1973ju,Gross:1974cs}.
Historically, these works date back to the very beginning of QCD itself \cite{Fritzsch:1973pi}.
In fact, objects equivalent to $\gamma_{ij}^{(0)}(N)$ were first encountered by Gribov and Lipatov~\cite{Gribov:1972ri}, who extracted the leading logarithms in QED-like ladder diagrams with kinematics corresponding to deep-inelastic scattering (DIS).
Perhaps, the most famous calculations of $\gamma_{ij}^{(0)}(N)$ were performed by Gross and Wilczek~\cite{Gross:1973ju,Gross:1974cs}, who directly treated twist-two Wilson operators.
Later, Altarelli and Parisi~\cite{Altarelli:1977zs} introduced the concept of spliting functions.
The picture was then completed by Dokshitzer \cite{Dokshitzer:1977sg}, who put everything in place.

The next-to-leading-order (NLO) \cite{Floratos:1977au,GonzalezArroyo:1979df,Floratos:1978ny,GonzalezArroyo:1979he,Gonzalez-Arroyo:1979kjx,Curci:1980uw,Furmanski:1980cm} and next-to-next-to-leading-order (NNLO) results \cite{Larin:1993vu,Larin:1996wd,Moch:2004pa,Vogt:2004mw,Ablinger:2014nga,Blumlein:2021enk,Blumlein:2022gpp,Gehrmann:2023ksf} are fully known analytically.
The NLO calculations were performed along the lines of Refs.~\cite{Altarelli:1977zs,Dokshitzer:1977sg,Gross:1973ju,Gross:1974cs} and the NNLO ones by solving recurrence relations for master integrals with arbitrary powers of propagators in the denominator or arbitrary powers of four-momentum in the numerator.

At next-to-next-to-next-to-leading order (NNNLO), only partial results are available so far \cite{Gracey:1994nn,Velizhanin:2011es,Velizhanin:2014fua,Davies:2016jie,Moch:2017uml,Davies:2017hyl,Moch:2018wjh,Das:2020adl,Moch:2021qrk,Falcioni:2023luc,Gehrmann:2023cqm,Falcioni:2023vqq,Falcioni:2023tzp,Moch:2023tdj,Gehrmann:2023iah,Falcioni:2024xyt,Falcioni:2024xav,Falcioni:2024qpd,Kniehl:2025ttz,Falcioni:2025hfz}.
This is partly due to the fact that the methods mentioned above face significant problems at NNNLO simply because the involved Feynman integrals are considerably more complicated.
In the case of the Mellin space method outlined above, the complexity of the four-loop computations rapidly grows with $N$, so that the quest for high-$N$ moments has come to a grinding halt.
Further progress is limited by the availability of more powerful computers.

In want of $\gamma_{ij}^{(3)}(N)$ for general value of $N$, we have to resort to approximations for $P_{ij}^{(3)}(x)$ \cite{Larin:1996wd,Moch:2017uml,Moch:2023tdj,Falcioni:2025hfz}, whose errors are difficult to control and bound to be particularly large at low values of $x$ (see, {\it e.g.}, Figs.~5 and 6 in Ref.~\cite{Moch:2017uml}).
This strongly motivates us to recover all-$N$ results for $\gamma_{ij}^{(3)}(N)$ as much as possible.

Recently, we found the all-$N$ result for the term proportional to $\zeta(3)$ of the flavor non-singlet anomalous dimension $\gamma_{\mathrm{ns}}^{(3)\pm}(N)$ at four loops \cite{Kniehl:2025jfs}, which is denoted as $\gamma_{\zeta_3}^{(3)}(N)$ in Eq.~\eqref{eq:zeta3} below.
In Ref.~\cite{Kniehl:2025jfs}, for lack of space, we presented this for the important case of QCD, with $n_c=3$, and also discussed the respective pieces $P_{\zeta_3}^{(3)\pm}(x)$ of the splitting functions $P_{\mathrm{ns}}^{(3)\pm}(x)$ in the limits $x\to0$ and $x\to1$.
The main purpose of the present paper is to provide the interested reader with full results for generic value of $n_c$, so as to display the decompositions in terms of color factors, both for $\gamma_{\zeta_3}^{(3)}(N)$ and $P_{\zeta_3}^{(3)\pm}(x)$.
After submission of Ref.~\cite{Kniehl:2025jfs}, we learned about an independent determination of the all-$N$ result for $\gamma_{\zeta_3}^{(3)}(N)$ \cite{Moch:2025pri}, which agrees with ours.

Besides the quark flavor asymmetry, we also present the all-$N$ result for the transcendental part of the valence anomalous dimension $\gamma_{\mathrm{ns}}^{(3)\mathrm{v}}(N)$ at four loops, which, to the best of our knowledge, has not been published elsewhere.
Also here, we find agreement with an independent calculation \cite{Moch:2025pri}.

We note in passing that a powerful new approach to directly deriving analytic all-$N$ results for anomalous dimensions at four loops has recently been proposed in Ref.~\cite{Gehrmann:2023iah} and successfully applied to the flavor non-singlet quark operators in Eq.~\eqref{eq:nonsin} in order to extract the QED-like $C_F^3n_f$ contribution.
This relies on the computation of off-shell operator matrix elements and employs integration-by-parts reductions and differential equations with respect to a tracing parameter.
It remains to be seen how this can be extended to more complicated color factors.

This paper is organized as follows.
In Section~\ref{sec:for}, we explain in detail the methodology of analytic reconstruction used here.
In Section~\ref{sec:N}, we restate our QCD result for $\gamma_{\zeta_3}^{(3)}(N)$ and examine its limits for $N\to0$ and $N\to\infty$.
In Section~\ref{sec:x}, we present our QCD results for $P_{\zeta_3}^{(3)\pm}(x)$ and investigate their limits for $x\to0$ and $x\to1$.
In Section~\ref{sec:val}, we list our all-$N$ result for the transcendental part of $\gamma_{\mathrm{ns}}^{(3)\mathrm{v}}(N)$ and the exact $x$ dependence of the corresponding contribution to $P_{\mathrm{ns}}^{(3)\mathrm{v}}(x)$ for SU($n_c$) color gauge group.
In Section~\ref{sec:dis}, we offer a numerical discussion and study the relative importance of the new $\gamma_{\zeta_3}^{(3)}(N)$ term relative to the rational and the other transcendental contributions to $\gamma_{\mathrm{ns}}^{(3)}(N)$.
Section~\ref{sec:con} contains a brief summary.
In Appendix~\ref{app:N}, we list $\gamma_{\zeta_3}^{(3)}(N)$ for SU($n_c$) color gauge group together with its $N\to0$ and $N\to\infty$ asymptotic expressions.
In Appendix~\ref{app:x}, we do the same for $P_{\zeta_3}^{(3)\pm}(x)$ and the limits $x\to0$ and $x\to1$.

\section{Formalism}
\label{sec:for}

In this section, we focus on $\gamma_{\mathrm{ns}}^{(3)\pm}(N)$.
From the operator product expansion it follows that $\gamma_{\mathrm{ns}}^{(3)\pm}(N)$ vanishes for odd/even values of $N$. It is therefore convenient to combine both cases into a single function, $\gamma_{\mathrm{ns}}^{(3)}(N)$.
This quantity can be decomposed into rational and transcendental contributions as
\begin{equation}
\gamma_{\mathrm{ns}}^{(3)}(N)=
\gamma_{\mathrm{rat}}^{(3)}(N)
+\zeta_3\gamma_{\zeta_3}^{(3)}(N)
+\zeta_4\gamma_{\zeta_4}^{(3)}(N)
+\zeta_5\gamma_{\zeta_5}^{(3)}(N)\,,
\label{eq:zeta3}
\end{equation}
where $\zeta_k=\zeta(k)$ denotes Riemann's zeta function.
A contribution proportional to $\zeta_2=\pi^2/6$ is excluded by the no-$\pi^2$ theorem \cite{Davies:2017hyl,Jamin:2017mul,Baikov:2018wgs}.

Detailed inspection of the available all-$N$ results for $\gamma_{ij}^{(n)}(N)$ at $(n+1)$ loops and parts thereof shows that the functional dependence is fully captured by nested harmonic sums (NHSs),
\begin{equation}
S_{m_1\ldots m_M}(N)=\sum^{N}_{i=1} \frac{[\sign(m_1)]^{i}}{i^{\vert m_1\vert}}
\,S_{m_2\ldots m_M}(i)\,,
\end{equation}
where $S(N)=1$ and $m_j\in\mathbb{Z},\ m_j\neq -1$, with weight $w=\sum_{j=1}^M|m_j|\le2n+1$, and their counterparts with arguments $N\pm1$.
For fixed weight $w$, the number of such sums is $3[(1-\sqrt{2})^w+(1+\sqrt{2})^w]/2$.
Therefore, these functions serve as a natural basis for an ansatz to analytically reconstruct the unknown $\gamma_{ij}^{(n)}(N)$ for arbitrary $N$.
Unfortunately, the dimension of this function space rapidly increases with $n$, taking the values $3, 21, 123, 717, \ldots$ for $n=0,1,2,3,\ldots$.

It is worth noting that $\zeta_k$ carries weight $w=k$, and each power of $n_f$ counts as weight one. When such factors appear overall, they effectively reduce the maximal weight of the basis functions required. This explains why the $\zeta_5$ \cite{Moch:2017uml}, $\zeta_4$ \cite{Davies:2017hyl}, $n_f^3$ \cite{Gracey:1994nn}, and $n_f^2$ \cite{Davies:2016jie} contributions to $\gamma_{\mathrm{ns}}^{(3)}(N)$ have long been known in closed form.
In the generic four-loop case ($n=3$), however, the number of possible basis elements typically far exceeds the number of constraints provided by the known low-$N$ moments of $\gamma_{ij}^{(n)}(N)$.

The key empirical observation is that the coefficients in the ansatz are typically simple rational numbers, involving moderate integers and powers of $2$ and $3$ in the denominators. This suggests that they can be determined using methods from number theory. Specifically, the system of linear equations derived from the known moments may be interpreted as a Diophantine system with fewer equations than unknowns. By applying the Lenstra--Lenstra--Lov\'{a}sz (LLL) algorithm \cite{Lenstra82factoringpolynomials}, as implemented in the program package \texttt{fplll} \cite{fplll}, to the corresponding matrix, one obtains a reduced matrix whose rows form solutions to the original system with minimal Euclidean norm.

This method for reconstructing the all-$N$ forms of anomalous dimensions from a finite number of Mellin moments was first proposed in the context of $\mathcal{N}=4$ supersymmetric Yang--Mills theory (SYM) \cite{Velizhanin:2010cm} and was then successfully applied in ${\mathcal N}=4$ \cite{Velizhanin:2013vla,Marboe:2014sya,Marboe:2016igj,Kniehl:2020rip,Kniehl:2021ysp,Velizhanin:2021bdh,Kniehl:2023bbk,Kniehl:2024tvd} and $\mathcal{N}=2$ SYM \cite{Kniehl:2023bbk}, and also in QCD \cite{Davies:2016jie,Moch:2017uml,Kniehl:2025ttz,Velizhanin:2012nm}.

Fortunately, the rapid growth of the function space dimensionality is drastically reduced in the flavor non-singlet sector through a generalized Gribov--Lipatov reciprocity relation~\cite{Gribov:1972rt}, originally observed as $P_{ii}^{(0)}(x)=-xP_{ii}^{(0)}(1/x)$ or, equivalently, as the quasi-invariance of $\gamma_{ii}^{(0)}(N)$ under $N\to-1-N$.
This symmetry implies that, at higher orders, the invariance no longer holds for the full $\gamma_{\mathrm{ns}}^{(n)}(N)$, but only for its reciprocity-respecting (RR) part $\mathcal{P}_{\mathrm{ns}}^{(n)}(N)$, which is determined through a self-tuning relation \cite{Dokshitzer:2005bf,Dokshitzer:2006nm,Basso:2006nk,Chen:2020uvt},
\begin{equation}
\gamma(N)=\mathcal{P}(N-\gamma(N)-\beta(a_s)/a_s)\,,
\end{equation}
or equivalently,
\begin{equation}
\mathcal{P}(N)=\gamma(N+\mathcal{P}(N)+\beta(a_s)/a_s)\,,
\label{eq:rr}
\end{equation}
where
\begin{equation}
\mu^2\frac{\mathrm{d}a_s}{\mathrm{d}\mu^2}=
\beta(a_s)=-\sum_{n=0}^\infty b_n\,a_s^{n+2}\,,
\end{equation}
is the QCD beta function.

By Taylor expanding Eq.~\eqref{eq:rr} and iteratively substituting it into itself, one can express $\mathcal{P}$ entirely in terms of $\gamma$ and its derivatives with respect to $N$,
\begin{equation}
\mathcal{P}=\gamma+\sum_{i=1}^\infty\frac{1}{(i+1)!}\,\frac{\mathrm{d}^i(\gamma+\beta/a_s)^{i+1}}{\mathrm{d}N^i}\,.
\label{eq:rr1}
\end{equation}
Upon perturbative expansion, one obtains
\begin{equation}
\mathcal{P}=a_s\gamma^{(0)}+\sum_{n=1}^\infty a_s^{n+1}(\gamma^{(n)}+\NP^{(n)})\,,
\end{equation}
with
\begin{eqnarray}
\NP^{(1)}&=&(b_{0} + \gamma^{(0)})\dot\gamma^{(0)}\,,\nonumber\\
\NP^{(2)}&=&(b_{1} + \gamma^{(1)})\dot\gamma^{(0)}
+ (b_{0} + \gamma^{(0)})[\dot\gamma^{(1)}+(\dot\gamma^{(0)})^2] 
+ \frac{1}{2}(b_{0}+\gamma^{(0)})^2 \ddot\gamma^{(0)}\,,
\nonumber\\
\NP^{(3)}&=&
(b_{2} + \gamma^{(2)})\dot\gamma^{(0)}
+(b_{1} + \gamma^{(1)})[\dot\gamma^{(1)}+(\dot\gamma^{(0)})^2]
+(b_{1} + \gamma^{(1)})(b_{0} + \gamma^{(0)})\ddot\gamma^{(0)}
\nonumber\\
&&{}+ (b_{0} + \gamma^{(0)})[\dot\gamma^{(2)}+2\dot\gamma^{(1)}\dot\gamma^{(0)}
  +(\dot\gamma^{(0)})^3]
+ \frac{1}{2}(b_0+\gamma^{(0)})^2(\ddot\gamma^{(1)}
+3\dot\gamma^{(0)}\ddot\gamma^{(0)})
\nonumber\\
&&{}+ \frac{1}{6}(b_{0} + \gamma^{(0)})^3 \dddot\gamma^{(0)}\,,
\label{eq:nrr}
\end{eqnarray}
where dots denote derivatives with respect to $N$, and so on.
In this way, the anomalous dimension is uniquely decomposed as $\gamma=\mathcal{P}-\NP$, where the non-RR part $\NP$ at each order is determined by lower-order anomalous dimensions and their derivatives. Consequently, the reconstruction of $\gamma_{\mathrm{ns}}^{(n)}(N)$ for all $N$ reduces to the corresponding problem for $\mathcal{P}_{\mathrm{ns}}^{(n)}(N)$.

This has the crucial advantage that the relevant function space collapses to the significantly smaller RR subspace spanned by binomial harmonic sums (BHSs),
\begin{equation}
\BS_{m_1\ldots m_M}(N)=(-1)^N\sum_{i=1}^{N}(-1)^i\binom{N}{i}\binom{N+i}{i}S_{m_1,...,m_M}(i)\,,
\end{equation}
where $m_j\in\mathbb{N}$. For fixed weight $w$, there are only $2^{w-1}$ such sums. 
Furthermore, although shifts in the arguments of harmonic sums give rise to the denominators $1/N$ and $1/(N+1)$, such terms occur only in the specific RR combination $\eta=1/[N(N+1)]$.

This naturally leads to the ansatz
\begin{equation}
\mathcal{P}_{\mathrm{ns}}^{(n)}(N)=\sum_{w=0}^{2n+1}c_w\BS_{\vec{\mathbf{m}}_w}(N)
+\sum_{k=1}^{2n+1}\sum_{w=0}^{2n+1-k}c_{kw}\eta^k\BS_{\vec{\mathbf{m}}_w}(N)\,,
\label{AnsatzOddEven}
\end{equation}
with rational coefficients $c_w$ and $c_{kw}$, possibly multiplied by factors $\zeta_k$.
However, this ansatz can be only applied to either even or odd values of $N$. A further observation makes it possible to unify the results for both even and odd $N$, while simultaneously reducing the size of the basis.
Specifically, the BHSs of highest weight appear with identical coefficients in the general expressions for both parity cases.
Therefore, we may write the ansatz in the following unified form:
\begin{equation}
\mathcal{P}_{\mathrm{ns}}^{(n)}(N)=c_{2n+1}\BS_{\vec{\mathbf{m}}_{2n+1}}(N)
+\sum_{w=0}^{2n}\varepsilon c_w^\varepsilon \BS_{\vec{\mathbf{m}}_w}(N)
+\sum_{k=1}^{2n+1}\sum_{w=0}^{2n+1-k}\varepsilon c_{kw}^\varepsilon\eta^k\BS_{\vec{\mathbf{m}}_w}(N)\,,
\label{AnsatzFull}
\end{equation}
with sign factor $\varepsilon = (-1)^N$, and use the available results for fixed $N$, for both even and odd values, improving the performance of the LLL algorithm \cite{Lenstra82factoringpolynomials,fplll}.

\boldmath
\section{QCD results in $N$ space}
\label{sec:N}
\unboldmath

The analytic reconstruction of $\gamma_{\zeta_3}^{(3)}(N)$ proceeds as explained in Section~\ref{sec:for}.
The fermionic contributions proportional to $n_f^3$, $n_f^2$, $n_f$, and the purely gluonic contribution, devoid of $n_f$, require BHSs of maximum weight $w=1,2,3,4$, respectively, namely $2\times3+1$ minus 3 from $\zeta_3$ minus $k$ from $n_f^k$.
The extraction of the $n_f^3$ and $n_f^2$ contributions is straightforward and reproduces the results of Refs.~\cite{Gracey:1994nn,Davies:2016jie}, respectively.
The residual color factors include $C_F^3n_f$, $C_F^2C_An_f$, $C_FC_A^2n_f$, $d_{44}^{\mathrm{RR}}n_f$, $C_F^4$, $C_F^3C_A$, $C_F^2C_A^2$, $C_FC_A^3$, $d_{44}^{\mathrm{RA}}$, where $d_{44}^{\mathrm{RR}}=d_F^{abcd}d_F^{abcd}/\nr=(n_c^2-1)(n_c^4-6n_c^2+18)/(96n_c^3)$ and $d_{44}^{\mathrm{RA}}=d_F^{abcd}d_A^{abcd}/\nr=(n_c^2-1)(n_c^2+6)/48$.
The basis for the $n_f$ contribution contains $26$ functions of weight $w=3$, which we extend by including the terms with weight $w=4$ that appear in the large-$n_c$ limit~\cite{Moch:2017uml}.
For the purely gluonic contribution, we consider the complete weight $w=4$ basis consisting of 54 terms.

The contributions proportional to the quartic color structures $d_{44}^{RA}$ and $d_{44}^{RR}$ appear for the first time at four loops and hence are RR by themselves.
Their reconstruction is straightforward, since the unknown coefficients are relatively simple numbers.
The $n_f$ contributions can likewise be reconstructed without difficulty due to the relatively small size of the basis.
By contrast, the purely gluonic contribution poses additional challenges.
The main difficulties stem from the possible appearance of higher powers of $2$ and $3$ in the denominators of the coefficients, as well as from rather large numerical values in the numerators accompanying the lower-weight terms.
To exclude the most problematic terms, we first solve a subset of the initial equations and then rescale selected terms by powers of $3$ and remove certain elements from the basis by trial and error.
This enables us to successfully apply the LLL algorithm \cite{Lenstra82factoringpolynomials,fplll}, to tackle the residual color structures.
Our all-$N$ result for $\gamma_{\zeta_3}^{(3)}(N)$ in SU($n_c$) color gauge theory is presented in Eq.~\eqref{eq:adN}.

The $C_F^3n_f$ \cite{Gehrmann:2023iah}, large-$n_c$ \cite{Moch:2017uml}, and $\mathcal{N}=4$ SYM \cite{Kniehl:2021ysp,Kniehl:2024tvd,Beisert:2006ez,Bajnok:2008qj}, results, not injected in our analytic reconstruction, serve as valuable cross checks for Eq.~\eqref{eq:adN}.
Specifically, the large-$n_c$ limit is reached by putting $2C_F=C_A=n_c$, $d_{44}^{\mathrm{RA}}=n_c^3/96$, and $d_{44}^{\mathrm{RA}}=n_c^4/48$, and the result of $\mathcal{N}=4$ SYM is extracted by putting $C_F=C_A=n_c$, $d_{44}^{\mathrm{RA}}=n_c^4/24$, $n_f=0$ and retaining only the BHSs with weight $w=4$.
Our result in Eq.~\eqref{eq:adN} passes all these checks.
Furthermore, it reproduces the moments for $N=17,\ldots,20$ \cite{Moch:2025pri}, unavailable for the analysis of Ref.~\cite{Kniehl:2025jfs}.

In QCD, with $n_c=3$, our all-$N$ result for $\gamma_{\zeta_3}^{(3)}(N)$ reads \cite{Kniehl:2025jfs}
{\small
\begin{eqnarray}
  \gamma_{\z3}^{(3)}&=&
\frac{1024}{27}
\Big[
-\frac{26307}{128}
+\frac{1}{144}(8111+20763\mm)\eta
+\frac{477}{2}\HS_{1}
+\frac{5}{8}(10-53\mm)\D1^2
+\frac{1}{48}(1721+5970\mm)\eta^2
\nonumber\\&&{}
-\frac{1}{48}(15977-2586\mm)\eta\HS_{1}
+\frac{25}{4}\HS_{2}
-\frac{1}{144}(31059-71\mm)\HS_{-2}
-\frac{185}{4}\mm\D1^3
+\frac{1}{288}(3691-2389\mm)\eta^3
\nonumber\\&&{}
-10\mm\D1^2\HS_{1}
-\frac{1}{48}(1963+810\mm)\eta^2\HS_{1}
-\frac{1}{24}(7289+528\mm)\eta\HS_{2}
-\frac{1}{144}(8928+2269\mm)\eta\HS_{-2}
\nonumber\\&&{}
+\frac{1}{12}(7289+408\mm)\eta\HS_{1,1}
+\frac{1}{8}(399+62\mm)\HS_{-3}
-\frac{1}{8}(812+31\mm)\HS_{3}
-\frac{1}{4}(351+62\mm)\HS_{-2,1}
\nonumber\\&&{}
+\frac{81}{2}\HS_{1,-2}
-15\mm\D1^4
+\frac{1}{24}(135-71\mm)\eta^4
-20\mm\D1^3\HS_{1}
+\frac{1}{48}(9217+1396\mm)\eta^3\HS_{1}
+10\D1^2\HS_{-2}
\nonumber\\&&{}
-\frac{1}{24}(603+532\mm)\eta^2\HS_{-2}
-\frac{1}{24}(6805-816\mm)\eta^2\HS_{1,1}
+\frac{1}{48}(6805-1056\mm)\eta^2\HS_{2}
-\frac{1}{8}(4+31\mm)\eta\HS_{3}
\nonumber\\&&{}
-\frac{1}{2}(214+31\mm)\eta\HS_{-2,1}
+\frac{1}{4}(120+31\mm)\eta\HS_{-3}
+27\eta\HS_{1,-2}
-53\HS_{-3,1}
+44\HS_{-2,2}
-60\HS_{1,-3}
+\HS_{1,3}
\nonumber\\&&{}
+37\HS_{2,-2}
-85\HS_{3,1}
-68\HS_{-2,1,1}
+214\HS_{1,-2,1}
-54\HS_{1,1,-2}
-\frac{7}{6}\HS_{-4}
+\frac{119}{6}\HS_{4}
+\frac{133}{3}\HS_{-2,-2}
\Big]
\nonumber\\&&{}
+\frac{512}{27}\nf
\Big[
\frac{15241}{96}
+\frac{1}{48}(5819-360\mm)\eta
-\frac{3463}{12}\HS_{1}
+\frac{5}{4}(25+2\mm)\D1^2
+\frac{1}{24}(382-219\mm)\eta^2
\nonumber\\&&{}
+\frac{1}{2}(46+7\mm)\eta\HS_{1}
+\frac{35}{4}\HS_{-2}
+\frac{45}{4}\HS_{2}
+5(4+\mm)\D1^3
+\frac{1}{12}(113-10\mm)\eta^3
+40\D1^2\HS_{1}
\nonumber\\&&{}
-\frac{1}{2}(52-7\mm)\eta^2\HS_{1}
+\frac{5}{2}\eta\HS_{-2}
+\frac{29}{2}\eta\HS_{2}
-69\eta\HS_{1,1}
-\frac{197}{6}\HS_{3}
+\HS_{-3}
-7\HS_{-2,1}
-5\HS_{1,-2}
\nonumber\\&&{}
+40\HS_{1,2}
+40\HS_{2,1}
-\frac{69}{2}\eta^2\HS_{2}
+69\eta^2\HS_{1,1}
-\frac{69}{2}\eta^3\HS_{1}
\Big]\,,
\label{eq:Nqcd}
\end{eqnarray}
}
where $D_1=1/(N+1)$ and the argument $N$ has been dropped.

The limits $N\to0$ and $N\to\infty$ of Eq.~\eqref{eq:Nqcd} are of special interest.
Using tools from Refs.~\cite{Velizhanin:2020avm,Velizhanin:2022faj} and an appropriately modified version of the \texttt{MATHEMATICA} code from Ref.~\cite{Velizhanin:2021bdh}, we obtain for $N\to0+\w$
\begin{eqnarray}
\gamma_{\z3}^{(3)+}(N)&=&
\frac{8192}{81\w^4}
+ \frac{64}{81\w^3}(206 \nf-295)
+ \frac{64}{81\w^2}[8320 + 10613 \z2-\nf (455 + 828 \z2)]
\nonumber\\&&{}+
\frac{32}{243 \w}[3 \nf (7279 + 3888 \z2 + 1656 \z3) - 4 (10523 + 51918 \z2 + 11532 \z3)]\,,
\nonumber\\
\gamma_{\z3}^{(3)-}(N)&=&
\frac{26368}{81\w^4}
+ \frac{128}{243\w^3}(369 \nf-952)
+ \frac{64}{81\w^2}[-3169+7821 \z2 - \nf (137 + 828 \z2)]
\nonumber\\&&{}+
\frac{32}{243\w}[4 (3181 - 36924 \z2 - 10692 \z3)+3 \nf (6727 + 3552 \z2 + 1656 \z3)]\,,
\end{eqnarray}
and for $N\to\infty$
\begin{eqnarray}
\gamma_{\z3}^{(3)}&=&
512 \z2 \Big(\ln^2N+\frac{\ln N}{N}\Big)
+ \frac{128}{81} [ 54 (106 - 9 \z2 - 36 \z3) - (3463 - 510 \z2)\nf] \ln N
\nonumber\\&&{}
+ \frac{8}{243} 
[-236763 
+4(32859-71\mm)\z2
- 102672 \z3 
- 53352 \z4
+ 6 (15241 + 660 \z2 
\nonumber\\&&{}
+ 1096 \z3) \nf]
+\frac{64}{81}[
54 (106 - 21 \z2 - 36 \z3)
- (3463 - 510 \z2) \nf
]\frac{1}{N}\,.
\end{eqnarray}

\boldmath
\section{QCD results in $x$ space}
\label{sec:x}
\unboldmath

The splitting functions $P_{ij}(x)$ are recovered from the anomalous dimensions $\gamma_{ij}(N)$ upon analytic continuation to the complex $N$ plane via inverse Mellin transformation,
\begin{equation}
P_{ij}(x)=-\frac{1}{2\pi i}\int_{c-i\infty}^{c+i\infty}dN\,x^{-N}\gamma_{ij}(N)\,,
\label{eq:inv}
\end{equation}
with a suitable real number $c>0$.
In our case, it is convenient to resort to the \texttt{harmpol} \cite{Remiddi:1999ew} package written in \texttt{FORM} \cite{Vermaseren:2000nd} language, which is particularly suited for NHSs.
Notice that inverse Mellin transformation of NHSs also generates multiple zeta values.
In fact, $P_{\zeta_3}^{(3)\pm}(x)$, which multiply $\zeta_3$, turn out to contain $\zeta_2$, $\zeta_3$, and $\zeta_4$.

Our new results for $P_{\zeta_3}^{(3)\pm}(x)$ in QCD read
\allowdisplaybreaks
\small
\begin{eqnarray}
P_{\z3}^{(3)+}&=&
\frac{512}{27}
\Big\{
\frac{1937}{36}(1-x)
+\frac{1}{72}(111270-56861x)\z2
+\frac{1}{12}(3412+3951x)\z3
-\frac{18611}{8}(1-x)\HPL_{1}
\nonumber\\&&{}
+\frac{1}{72}\big[9916+172645x+3(7301+7371x)\z2\big]\HPL_{0}
-31(1+x)\z2\HPL_{-1}
+\frac{1}{72}(4323-27418x)\HPL_{0,0}
\nonumber\\&&{}
-\frac{1}{12}(18545-8587x)\HPL_{0,1}
-2281(1-x)\HPL_{1,1}
-\frac{2261}{2}(1-x)\HPL_{1,0}
-\frac{5339}{72}(1+x)\HPL_{-1,0}
\nonumber\\&&{}
+\frac{31}{2}(1+x)\HPL_{-1,0,0}
+\frac{31}{4}(1-x)\HPL_{1,0,0}
-\frac{53}{4}(1-x)\HPL_{0,-1,0}
-\frac{6805}{24}(1+x)(2\HPL_{0,1,1}+\HPL_{0,1,0})
\nonumber\\&&{}
+31(1+x)\HPL_{-1,0,1}
+\frac{1}{6}(144+229x)\HPL_{0,0,0}
-\frac{1}{24}(7301+7053x)\HPL_{0,0,1}
+\frac{\pqqp}{24}
\Big(
5724
\nonumber\\&&{}
-486\z2
-1524\z3
-150\HPL_{0}
+1952\z2\HPL_{0}
+648\z2\HPL_{1}
-2529\HPL_{0,0}
-1064\HPL_{0,-1,0}
-476\HPL_{0,0,0}
\nonumber\\&&{}
-2040\HPL_{0,0,1}
+24\HPL_{1,0,0}
\Big)
-\frac{\pqqm}{72}
\Big(
5976\z2
-7164\z3
+13464\z2\HPL_{-1}
-15494\HPL_{0}
-4080\z2\HPL_{0}
\nonumber\\&&{}
-2916\HPL_{-1,0}
-4149\HPL_{0,0}
-7434\HPL_{0,1}
-84\HPL_{0,0,0}
+3816\HPL_{0,0,1}
-3888\HPL_{-1,-1,0}
-4320\HPL_{-1,0,0}
\nonumber\\&&{}
-15408\HPL_{-1,0,1}
+2664\HPL_{0,-1,0}
-3168\HPL_{0,1,0}
-4896\HPL_{0,1,1}
\Big)
+\frac{1}{576}(236763-131152\z2
\nonumber\\&&{}
+102672\z3+92232\z4)\delta(1-x)
\Big\}
+\frac{256}{9}\nf
\Big[
-\frac{353}{72}(1-x)
-(65-73x)\z2
-23(1+x)\z3
\nonumber\\&&{}
-\frac{1}{36}(215+2927x)\HPL_{0}
+\frac{271}{3}(1-x)\HPL_{1}
-23(1+x)\z2\HPL_{0}
+\frac{4}{9}(11+10x)\HPL_{0,0}
+69(1-x)\HPL_{1,0}
\nonumber\\&&{}
+(65-73x)\HPL_{0,1}
+138(1-x)\HPL_{1,1}
+23(1+x)\HPL_{0,0,1}
+23(1+x)\HPL_{0,1,0}
+46(1+x)\HPL_{0,1,1}
\nonumber\\&&{}
-\frac{\pqqm}{12}
\Big(
38\z2
+35\HPL_{0}
+20\HPL_{-1,0}
-4\HPL_{0,0}
-28\HPL_{0,1}
\Big)
+\frac{\pqqp}{36}
(
510\z2
-3463
-135\HPL_{0}
\nonumber\\&&{}
-394\HPL_{0,0}
-480\HPL_{0,1}
-480\HPL_{1,0}
)
-\frac{1}{144}(15241+660\z2+1096\z3)\delta(1-x)
\Big]
\,,
\nonumber\\
P_{\z3}^{(3)-}&=&\frac{512}{27}\Big\{
-\frac{11767}{36}(1-x)
+\frac{1}{72}(65755-106818x)\z2
+\frac{1}{12}(3951+3412x)\z3
-31(1+x)\z2\HPL_{-1}
\nonumber\\&&{}
+\left[\frac{1}{72}(6508+103285 x)+\frac{1}{24}(7371+7301x)\z2\right]\HPL_{0}
-\frac{10223}{8}(1-x)\HPL_{1}
-\frac{2891}{72}(1+x)\HPL_{-1,0}
\nonumber\\&&{}
+\frac{1}{72}(11966+8571 x)\HPL_{0,0}
-\frac{1}{12}(11441-17803 x)\HPL_{0,1}
-\frac{2437}{2}(1-x)\HPL_{1,0}
-2417(1-x)\HPL_{1,1}
\nonumber\\&&{}
+\frac{31}{2}(1+x)\HPL_{-1,0,0}
+ 31(1+x)\HPL_{-1,0,1}
+\frac{159}{12}(1-x)\HPL_{0,-1,0}
+\frac{1}{6}(229+144 x)\HPL_{0,0,0}
\nonumber\\&&{}
-\frac{1}{24}(7053+7301x)\HPL_{0,0,1}
-\frac{6805}{24}(1+x)\HPL_{0,1,0}
-\frac{6805}{12}(1+x)\HPL_{0,1,1}
-\frac{31}{4}(1-x)\HPL_{1,0,0}
\nonumber\\&&{}
+\frac{\pqqm}{72}[3744\z2
-7164\z3
+13464\z2\HPL_{-1}
-(15565+4080\z2)\HPL_{0}
-2916\HPL_{-1,0}
-3033\HPL_{0,0}
\nonumber\\&&{}
-5202\HPL_{0,1}
-3888\HPL_{-1,-1,0}
-4320\HPL_{-1,0,0}
-15408\HPL_{-1,0,1}
+2664\HPL_{0,-1,0}
-84\HPL_{0,0,0}
\nonumber\\&&{}
+3816\HPL_{0,0,1}
-3168\HPL_{0,1,0}
-4896\HPL_{0,1,1}
]
+\frac{\pqqp}{24}[
5724
-1524\z3
-486\z2
+648\z2\HPL_{1}
\nonumber\\&&{}
-(150-1952\z2)\HPL_{0}
-2343\HPL_{0,0}
-1064\HPL_{0,-1,0}
-476\HPL_{0,0,0}
-2040\HPL_{0,0,1}
+24\HPL_{1,0,0}
]
\nonumber\\&&{}
+\frac{1}{576}(236763-131720\z2+102672\z3+92232\z4)\delta(1-x)
\Big\}
+\frac{256}{9}\nf\Big\{
\frac{199}{72}(1-x)
\nonumber\\&&{}
-\frac{1}{3}(195-219x)\z2
-23(1+x)\z3
-\frac{1}{36}(107+2771x+828(1+x)\z2)\HPL_{0}
+\frac{257}{3}(1-x)\HPL_{1}
\nonumber\\&&{}
+\frac{4}{9}(10+11x)\HPL_{0,0}
+\frac{1}{3}(195-219x)\HPL_{0,1}
+69(1-x)\HPL_{1,0}
+138(1-x)\HPL_{1,1}
+23(1+x)\HPL_{0,0,1}
\nonumber\\&&{}
+23(1+x)\HPL_{0,1,0}
+46(1+x)\HPL_{0,1,1}
+\frac{\pqqm}{12}(
38\z2
+35\HPL_{0}
+20\HPL_{-1,0}
-4\HPL_{0,0}
-28\HPL_{0,1}
)
\nonumber\\&&{}
+\frac{\pqqp}{36}(
510\z2
-3463
-135\HPL_{0}
-394\HPL_{0,0}
-480\HPL_{0,1}
-480\HPL_{1,0}
)
\nonumber\\&&{}
-\frac{1}{144}(15241+660\z2+1096\z3)\delta(1-x)
\Big\}\,,
\label{eq:x}
\end{eqnarray}
\normalsize
where $\HP_{\bm{\vec{a}}}(x)$ are harmonic polylogarithms (HPLs) and $\pqqp$ is the LO spliting function
\begin{equation}
\pqqp=\frac{2}{1-x}-1-x\,.
\end{equation}
For the ease of notation, we have dropped the argument $x$ of the HPLs in Eq.~\eqref{eq:x}.

For the reader's convenience, we recall the basic definition of the HPLs \cite{Remiddi:1999ew}.
The lowest-weight functions $H_m(x)$, with $w = 1$, are given by
\begin{equation}
  H_0(x)       =\ln x\,, \qquad
  H_{\pm 1}(x) = \mp \ln (1 \mp x)\,.
\end{equation}
The higher-weight functions, with $w \geq 2$, are recursively defined as
\begin{equation}
  H_{m_1,\ldots,m_w}(x) = 
  \left\{ 
    \begin{array}{ll}
      \displaystyle{\frac{1}{w!}\ln^w x\,,}
      & 
      \mbox{ if } m_1=\cdots=m_w=0\,, 
      \\[2ex]
      \displaystyle{ 
        \int_0^x dz\, f_{m_1}(z)H_{m_2,\dots,m_w}(z)\,,} 
      & 
      \mbox{ otherwise}\,,
    \end{array} \right.
\end{equation}
where
\begin{equation}
  f_0(x)       = x^{-1} \,, \qquad
  f_{\pm 1}(x) =  (1 \mp x)^{-1} \,.
\end{equation}

The soft and hard limits, $x\to0$ and $x\to1$, of the splitting functions are of special interest.
They correspond to the limits $N\to0$ and $N\to\infty$ of the anomalous dimensions.
For $x\to0$, we have
\begin{eqnarray}
P_{\zeta_3}^{(3)+}(x)&=&
\frac{4096}{243}\ln^3x
+ \left(\frac{9440}{81}-\frac{6592}{81}n_f\right) \ln^2x
+\frac{64}{81}\left[ 8320+ 10613 \zeta_2-(455 + 828 \zeta_2)n_f\right]
\nonumber\\
&&{}\times\ln x+\frac{128}{243}(10523+ 51918 \zeta_2 + 11532 \zeta_3) 
-\frac{32}{81} (7279 + 3888 \zeta_2 + 1656 \zeta_3)n_f\,,
\nonumber\\
P_{\z3}^{(3)-}(x)&=&
\frac{13184}{243} \ln^3x
+ \left(\frac{60928}{243}-\frac{2624}{27}\nf\right) \ln^2x 
- \frac{64}{81}[3169 - 7963 \z2 + (137 + 828 \z2)\nf]
\nonumber\\&&{}\times\ln x
+\frac{64}{243}\Big(-6362 + 68041 \z2 + 11970 \z3\Big)
-\frac{32}{81}\Big(6727 + 3432 \z2 + 1656 \z3\Big)\nf\,,
\nonumber\\
\end{eqnarray}
and, for $x\to1$, we have
\begin{eqnarray}
P_{\zeta_3}^{(3)+}(x)&=&
  - 1024 \zeta_2 \left[\frac{\ln(1 - x)}{1 - x}\right]_+
+ \left[\frac{27136}{3}
-768 \zeta_2  
- 3072 \zeta_3 +n_f \left(-\frac{443264}{81} + \frac{21760}{27}\zeta_2\right)
\right]
\nonumber\\&&{}
\times\left[\frac{1}{1 - x}\right]_+
+ \left[\frac{23384}{3} 
-\frac{1049216}{243}\zeta_2
+ \frac{91264}{27}\zeta_3 
+ \frac{27328}{9}\zeta_4 \right.
\nonumber\\&&{}
-\left. n_f \left(\frac{243856}{81}
+\frac{3520}{27}\zeta_2
+ \frac{17536}{81}\zeta_3 \right) 
\right]\delta(1-x)
+ 1024 \zeta_2 \ln(1-x)
\nonumber\\&&{}
-\frac{244224}{27} 
+ \frac{20736}{27}\zeta_2
+ 3072\zeta_3
+ n_f\left(\frac{443264}{81}
- \frac{21760}{27}\zeta_2\right)\,,
\nonumber\\
P_{\zeta_3}^{(3)-}(x)&=&P_{\zeta_3}^{(3)+}(x)-\frac{4544}{243}\z2\delta(1-x)\,,
\label{eq:one}
\end{eqnarray}
where plus distributions $[d(x)]_+$ are defined as $\int_0^1\mathrm{d}x\,[d(x)]_+f(x)=\int_0^1\mathrm{d}x\,d(x)[f(x)-f(1)]$.
Notice that the extraordinary $[\ln(1-x)/(1-x)]_+$ terms in Eq.~\eqref{eq:one} will cancel against similar terms in $P_{\mathrm{rat}}^{(3)\pm}(x)$.

\section{Flavor non-singlet valence}
\label{sec:val}

To complete our discussion of the quark flavor non-singlet sector, we now address the valence anomalous dimensions $\gamma_{\mathrm{ns}}^{(n)\mathrm{v}}(N)$, which are decomposed as \cite{Moch:2004pa,Moch:2017uml}
\begin{equation}
\gamma_{\mathrm{ns}}^{(n)\mathrm{v}}(N)=\gamma_{\mathrm{ns}}^{(n)-}(N)+\gamma_{\mathrm{ns}}^{(n)\mathrm{s}}(N)\,,
\end{equation}
where $\gamma_{\mathrm{ns}}^{(n)\mathrm{s}}(N)$ are flavor-independent (``sea'') contributions.
The latter appear for the first time at three-loop order, for $n=2$, and are nonzero only for odd values of $N$ \cite{Larin:1993vu,Moch:2004pa}.
We recall that $\gamma_{\mathrm{ns}}^{(3)-}(N)$ emerges from Eqs.~\eqref{eq:Nqcd} and \eqref{eq:adN} for $n_c=3$ and arbitrary value of $n_c$, respectively, by retaining only odd values of $N$.

At four loops, $\gamma_{\mathrm{ns}}^{(3)\mathrm{s}}(N)$ are available for $N=3,5,\ldots,15$ \cite{Moch:2017uml}, the moment for $N=1$ being equal to zero.
In contrast to Eq.~\eqref{eq:zeta3}, there is no $\zeta_4$ term,
\begin{equation}
\gamma_{\mathrm{ns}}^{(3)\mathrm{s}}(N)=
\gamma_{\mathrm{rat}}^{(3)\mathrm{s}}(N)
+\zeta_3\gamma_{\zeta_3}^{(3)\mathrm{s}}(N)
+\zeta_5\gamma_{\zeta_5}^{(3)\mathrm{s}}(N)\,.
\label{eq:zeta3v}
\end{equation}
The color structures include $d_{33}n_f^2$, $d_{33}C_Fn_f$, and $d_{33}C_An_f$, where $d_{33}=d_{abc}d^{abc}/N_R=(n_c^2-1)(n_c^2-4)/(16n_c^2)$ \cite{Moch:2017uml}.
The coefficient of $d_{33}n_f^2$ in $\gamma_{\mathrm{ns}}^{(3)\mathrm{s}}(N)$ is rational and known for all values of $N$ \cite{Davies:2016jie}; it reaches through weight $w=6$, one unity beyond what is expected from the transcendentality counting rules explained in Section~\ref{sec:for}.

We wish to derive from the available moments \cite{Moch:2017uml} the all-$N$ expressions for $\gamma_{\zeta_3}^{(3)\mathrm{s}}(N)$ and $\gamma_{\zeta_5}^{(3)\mathrm{s}}(N)$, making up the transcendental part in Eq.~\eqref{eq:zeta3v}.
$\gamma_{\zeta_5}^{(3)\mathrm{s}}(N)$ only involves $d_{33}C_An_f$, while $\gamma_{\zeta_3}^{(3)\mathrm{s}}(N)$ also involves $d_{33}C_Fn_f$ \cite{Moch:2017uml}.
Furthermore, the known all-$N$ expression for $\gamma_{\mathrm{ns}}^{(2)\mathrm{s}}(N)$ \cite{Moch:2004pa} is devoid of $\zeta_k$ values.
That is $\zeta_3$ and $\zeta_5$ only enter at four loops, which implies that $\gamma_{\zeta_3}^{(3)\mathrm{s}}(N)$ and $\gamma_{\zeta_5}^{(3)\mathrm{s}}(N)$ are RR by themselves.
From the all-$N$ expression for $\gamma_{\mathrm{ns}}^{(2)\mathrm{s}}(N)$ \cite{Moch:2004pa}, we infer that the relevant function basis is likely to be constructed from BHSs, $\eta$, and $\nu=1/(N-1)-1/(N+2)$, with the BHSs being dressed by powers of $\eta$ or $\nu$.
Similarly to $\eta$, $\nu$ is RR and assigned weight $w=1$.
According to the above transcendentality counting rules, the basis functions entering $\gamma_{\zeta_3}^{(3)\mathrm{s}}(N)$ and $\gamma_{\zeta_5}^{(3)\mathrm{s}}(N)$ are expected to have maximum weights $w=3$ and $w=1$, respectively.

In the latter case, however, the $w=1$ hypothesis is found to fail, so that we proceed to $w=2$, where the most general function basis reads
\begin{equation}
  \{1,\eta,\nu,\BS_{1},\eta^2,\nu^2,\eta\BS_{1},\nu\BS_{1},\BS_2,\BS_{1,1}\}\,.
  \label{AgVz5}
\end{equation}
Analytic reconstruction by means of the LLL algorithm \cite{Lenstra82factoringpolynomials,fplll} is straightforward and yields
\begin{equation}
\gamma_{\zeta_5}^{(3)\mathrm{s}}(N)=d_{33}C_An_f\frac{160}{3}(1-8\eta+12\eta^2)\,.
\label{eq:vz5}
\end{equation}
We note that the $\mathcal{O}(1)$ term in Eq.~\eqref{eq:vz5} has to be canceled by other contributions from $\gamma^{(3)\mathrm{s}}(N)$ in order for the latter to exhibit the well-known asymptotic large-$N$ behavior being of $\mathcal{O}(1/N)$ \cite{Moch:2004pa,Dokshitzer:2005bf,Korchemsky:1988si,Albino:2000cp}.

We now turn to $\gamma_{\zeta_3}^{(3)\mathrm{s}}(N)$.
Our observations regarding the all-$N$ expression for $\gamma_{\mathrm{ns}}^{(2)\mathrm{s}}(N)$ \cite{Moch:2004pa} made above lead us to try the following ansatz of function basis:
\begin{eqnarray}
&&\{\eta,\nu,\eta^2,\nu^2,\eta\BS_{1},\nu\BS_{1},\eta^3,\nu^3,\eta^2\BS_{1},\nu^2\BS_{1},\eta\BS_{2},\eta\BS_{1,1},\nu\BS_{2},\nu\BS_{1,1},\eta^4,\nu^4,\eta^3\BS_{1},\nu^3\BS_{1},\eta^2\BS_{2},
  \nonumber\\
&&{}\qquad\eta^2\BS_{1,1},\nu^2\BS_{2},\nu^2\BS_{1,1},\eta\BS_{3},\eta\BS_{2,1},\eta\BS_{1,2},\eta\BS_{1,1,1},\nu\BS_{3},\nu\BS_{2,1},\nu\BS_{1,2},\nu\BS_{1,1,1}\}\,.
\label{AgVz3}
\end{eqnarray}
This allows us to analytically reconstruct the $d_{33}C_Fn_f$ term of $\gamma_{\zeta_3}^{(3)\mathrm{s}}(N)$ via the LLL algorithm \cite{Lenstra82factoringpolynomials,fplll}, but fails for the $d_{33}C_An_f$ term.
However, we do succeed also in the latter case if we join Eqs.~\eqref{AgVz5} and \eqref{AgVz3}.
Altogether, we have
\begin{eqnarray}
\gamma_{\zeta_3}^{(3)\mathrm{s}}(N)&=&
d_{33}C_Fn_f 128[
- 54\eta 
+ 16\nu 
- 82\eta^2 
- 60\eta^3 
- 16\eta^4 
+ \BS_1 ( 
49\eta 
- 16\nu
+ 58\eta^2 
+ 24\eta^3 
)
\nonumber\\
&&{}+4\BS_{1,1} 
(
- 5\eta 
+ 2\nu 
- 4\eta^2 
)
]
+d_{33}C_An_f
\frac{64}{3}[
121\eta - 32\nu + 168\eta^2 + 260\eta^3 + 96\eta^4
\nonumber\\
&&{}+ \BS_1 (
- 105\eta 
+ 32\nu 
- 165\eta^2 
- 78\eta^3 
)
 -\BS_2 ( 
1 
- 8\eta 
+ 12\eta^2 
)
\nonumber\\
&&{}+ \BS_{1,1} ( 
45\eta 
- 16\nu
+ 42\eta^2 
)
]\nonumber\\
&=&d_{33}C_Fn_f 128\big[
- 54\eta 
+ 16\nu 
- 82\eta^2 
- 60\eta^3 
- 16\eta^4 
\nonumber\\
&&{}
+ 2\HS_{1} ( 
49\eta 
-16\nu
+ 58\eta^2 
+ 24\eta^3 
)
-8(2\HS_{1,1} - \HS_{2})
(
5\eta 
- 2\nu 
+ 4\eta^2 
)
\big]
\nonumber\\
&&{}
+d_{33}C_An_f
\frac{64}{3}\big[
121\eta - 32\nu + 168\eta^2 + 260\eta^3 + 96\eta^4
 +2\HS_{-2} ( 
1 
- 8\eta 
+ 12\eta^2 
)
\nonumber\\
&&{}
- 2\HS_{1} (
105\eta 
- 32\nu 
+ 165\eta^2 
+ 78\eta^3 
)
+ 2(2\HS_{1,1} - \HS_{2}) ( 
45\eta 
- 16\nu
+ 42\eta^2 
)
\big]\,,
\label{eq:vz3}
\end{eqnarray}
where we have converted BHSs to NHSs in the second equality. 

It is interesting to observe that Eq.~\eqref{eq:vz5} conspires with the term proportional to $d_{33}C_An_f\HS_{-2}$ in the second equality of Eq.~\eqref{eq:vz3} to produce the first two terms of the well-known factor
\begin{equation}
f(N) = 5 \z5 
+ 4 \z3 \HS_{-2} 
+ 2 \HS_{5}
- 2 \HS_{-5} 
- 4 \HS_{4, 1} 
+ 4 \HS_{3, -2} 
- 4 \HS_{-2, -3} 
+ 8 \HS_{-2, -2, 1} \,,
\label{eq:fN}
\end{equation}
which was first encountered in three-loop coefficient functions of inclusive DIS~\cite{Vermaseren:2005qc}, later recovered in wrapping corrections to twist-two operators at four loops in $\mathcal{N}=4$ SYM \cite{Bajnok:2008qj}, and more recently considered in the discussion of terms with quartic color factors in the anomalous dimensions $\gamma_{gg}^{(3)}(N)$ at four loops in SU($n_c$) color gauge theory \cite{Moch:2018wjh}.
We thus conjecture that the as-yet unknown all-$N$ expression for $\gamma_{\mathrm{rat}}^{(3)\mathrm{s}}(N)$ in Eq.~\eqref{eq:zeta3v} contains the term $d_{33}C_An_f(32/3)(2\HS_{5}-2\HS_{-5}-4\HS_{4, 1}+4\HS_{3, -2}-4\HS_{-2, -3}+8\HS_{-2, -2, 1})(1-8\eta+12\eta^2)$.
As for the leading large-$N$ term, devoid of $\eta$, this is motivated by the necessity to compensate the $\mathcal{O}(1)$ term in Eq.~\eqref{eq:vz5} in the large-$N$ limit, which works because $f(N)=\mathcal{O}(1/N^2)$.
The inclusion of the subleading $\eta^{1,2}$ terms in the conjecture is just because of the intriguing appearance of the very same factor $(1-8\eta+12\eta^2)$ in Eqs.~\eqref{eq:vz5} and \eqref{eq:vz3}.

For $N\to0+\w$, we have
\begin{eqnarray}
  \gamma_{\zeta_5}^{(3)\mathrm{s}}(N)&=&
 d_{33}C_A n_f\frac{640}{3}\left(\frac{3}{\w^2}-\frac{8}{\w}\right)
\,,  \nonumber\\
\gamma_{\zeta_3}^{(3)\mathrm{s}}(N)&=&d_{33}n_f\Big\{
\frac{2048}{\w^4} (- \cf+\ca )
+\frac{256}{3\w^3}(6 \cf - 31 \ca)
+\frac{256}{\w^2} [(-31 + 24 \z2)\cf
  \nonumber\\&&
 + (29 - 15 \z2)\ca]
+\frac{64}{3\w} [12 (35 - 14 \z2 - 24 \z3)\cf - (575 - 202 \z2 - 192 \z3) \ca]
\Big\}
\,,\nonumber\\
\end{eqnarray}
and, for $N\to\infty$, we have
\begin{eqnarray}
  \gamma_{\zeta_5}^{(3)\mathrm{s}}(N)&=&
 d_{33}C_A n_f\frac{160}{3} 
\,,  \nonumber\\
\gamma_{\zeta_3}^{(3)\mathrm{s}}(N)&=&
-d_{33}n_f\ca
\frac{64}{3}  \z2
\,.
\end{eqnarray}

Applying the \texttt{harmpol} package \cite{Remiddi:1999ew} to Eqs.~\eqref{eq:vz5} and \eqref{eq:vz3}, we obtain
\begin{eqnarray}
P_{\zeta_5}^{(3)\mathrm{s}}(x)&=&
\frac{160}{3} \ca d_{33} \nf\Big[32(1-x) + 12(1 + x)\HP_{0} - \delta(1 - x)\Big]
\,,\label{eq:pvz5}\\
P_{\zeta_3}^{(3)\mathrm{s}}(x)&=&
-256 \cf d_{33} \nf\Big\{
43(1 - x) 
- 2(7 - 5x - 8x^2)\z2 
- 8(3 + x)\z3 
+ 61(1 - x)\HP_{1} 
\nonumber\\&&
+ [23 - 38x - 8(3 + x)\z2]\HP_{0} 
+ 2(1 - 2x)(1 + 2x)\HP_{0, 0} 
+ 2(7 - 5x - 8x^2)\HP_{0, 1} 
\nonumber\\&&
+ 4\left(\frac{1}{x} - 1\right)(2 + x)(1 + 2x)(\HP_{1, 0}+2\HP_{1, 1})
+ 8\HP_{0, 0, 0} 
+ 8(3 + x)\HP_{0, 0, 1} 
\nonumber\\&&
+ 16(1 + x)(\HP_{0, 1, 0} +2\HP_{0, 1, 1})
\Big\}
- \frac{64}{3}\ca d_{33} \nf\Big\{
-607(1-x) 
+ 24(8 - x)\z3 
\nonumber\\&&
+ \z2[202 - 18x - 64x^2 - \delta(1 - x)] 
+ 2\left[-157+ 53x +6(15 + x)\z2 - \frac{1}{1 + x}\right]\HP_{0} 
\nonumber\\&&
-422(1-x)\HP_{1} 
+ 64(1 + x)\HP_{-1, 0} 
- 4(31 - 8x^2)\HP_{0, 0} 
- 2(69 - 9x - 32x^2)\HP_{0, 1} 
\nonumber\\&&
+ 2\left(1-\frac{1}{x} \right)(16 + 55x + 16x^2)(\HP_{1, 0}+2\HP_{1, 1})
+24(1-x)\HP_{0, -1, 0} 
- 96\HP_{0, 0, 0} 
\nonumber\\&&
- 12(13 + x)\HP_{0, 0, 1} 
- 84(1 + x)(\HP_{0, 1, 0} +2\HP_{0, 1, 1})
\Big\}
\label{eq:pvz3}\,.
\end{eqnarray}

For $x\to0$, we have
\begin{eqnarray}
P_{\zeta_5}^{(3)\mathrm{s}}(x)&=&
 d_{33}C_A n_f\frac{640}{3}\left(3\ln x+8\right)
\,,\nonumber\\
P_{\zeta_3}^{(3)\mathrm{s}}(x)&=&
\frac{64}{3}d_{33}n_f\Big\{
16(- \cf+\ca )\ln^3x
-2(6 \cf - 31 \ca)\ln^2x
+12[(-31 + 24 \z2)\cf
\nonumber\\&&
+(29 - 15 \z2)\ca]\ln x
-12 (35 - 14 \z2 - 24 \z3)\cf + (575 - 202 \z2 - 192 \z3) \ca
\Big\}
\,,\nonumber\\
\end{eqnarray}
and, for $x\to1$, we have
\begin{eqnarray}
P_{\zeta_5}^{(3)\mathrm{s}}(x)&=&
- d_{33}C_A n_f\frac{160}{3} \delta(1-x)
\,,\label{eq:sz5}\\
P_{\zeta_3}^{(3)\mathrm{s}}(x)&=&
d_{33}n_f\ca\frac{64}{3}  \z2\delta(1-x)
\,.
\end{eqnarray}
We note that the contribution in Eq.~\eqref{eq:sz5} cancels against a corresponding term in $P_{\mathrm{rat}}^{(3)\mathrm{s}}(x)$.

\section{Discussion}
\label{sec:dis}

We now assess the phenomenological relevance of our new QCD results for $\gamma_{\zeta_3}^{(3)}(N)$ and $P_{\zeta_3}^{(3)\pm}(x)$ presented in Sections~\ref{sec:N} and \ref{sec:x}, respectively.
We do this by comparing the sizes of the individual contributions to $\gamma_{\mathrm{ns}}^{(3)}(N)$ in Eq.~\eqref{eq:zeta3}.
In the case of $\gamma_{\zeta_3}^{(3)}(N)$, we further distinguish between the new terms devoid of and linear in $n_f$, listed in Eq.~\eqref{eq:Nqcd}, and the terms quadratic and cubic in $n_f$, previously available from Refs.~\cite{Gracey:1994nn,Davies:2016jie}.
For our purposes, it is sufficient to rely on the moments for $N=1,\ldots,16$ \cite{Velizhanin:2011es,Velizhanin:2014fua,Moch:2017uml}.
The findings will then naturally carry over to $P_{\zeta_3}^{(3)\pm}(x)$ via Eq.~\eqref{eq:inv}.

\begin{figure}
\begin{center}
\begin{tabular}{cc}
  \includegraphics[width=0.47\textwidth]{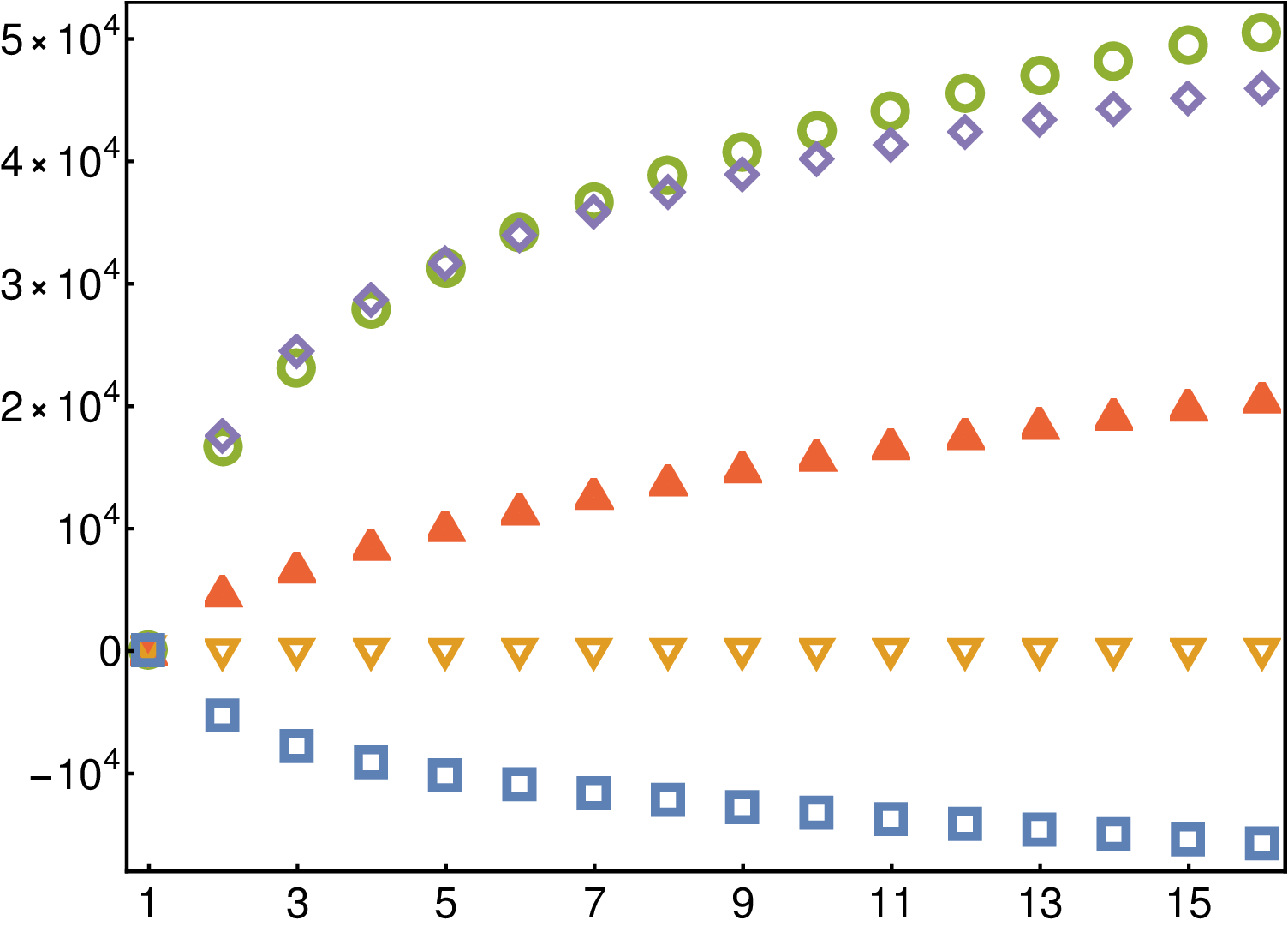}
  &
  \includegraphics[width=0.47\textwidth]{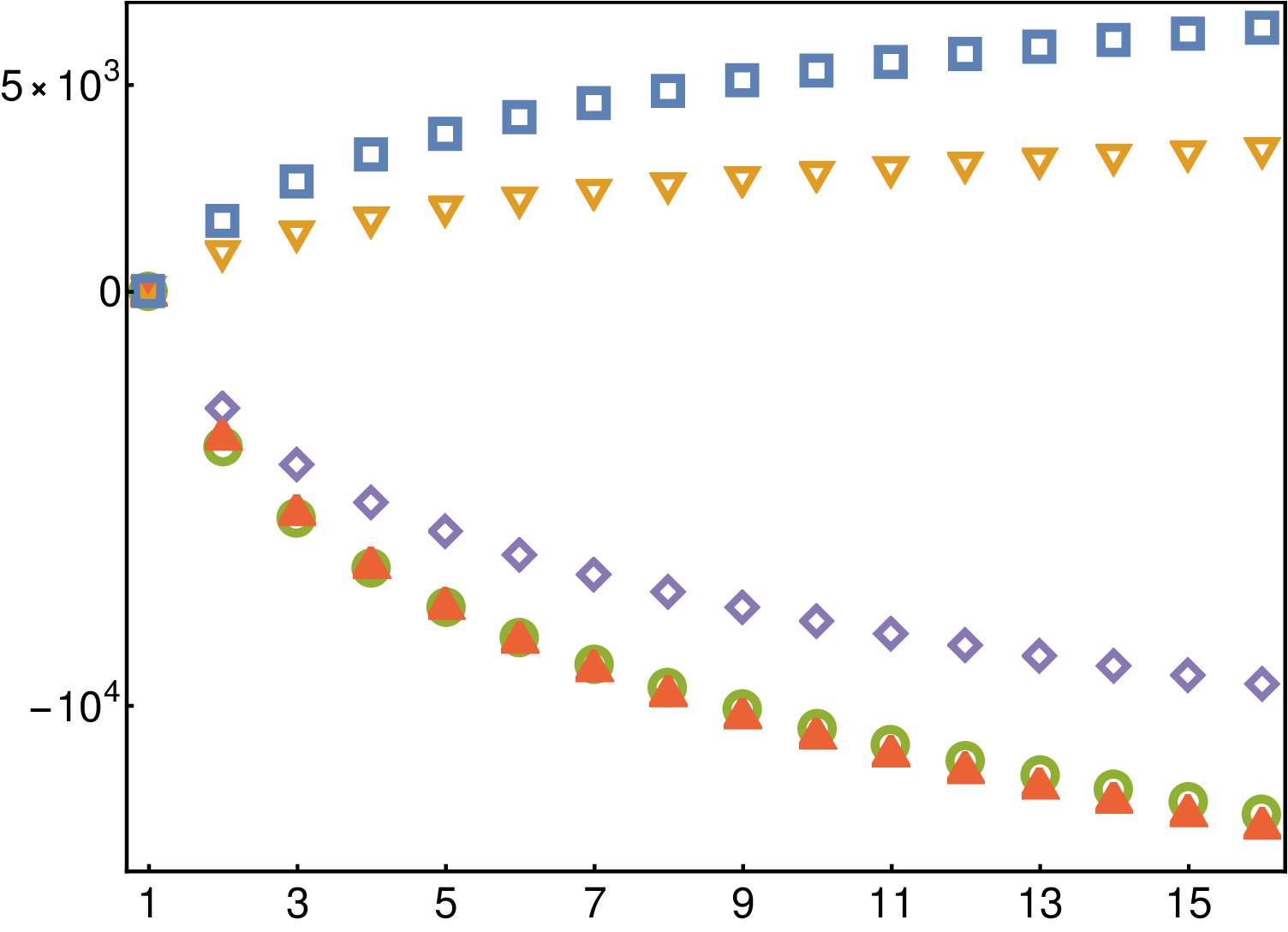}\\
(a) & (b)
\end{tabular}  
\caption{The (a) $n_f$-independent term and $(b)$ coefficient of $n_f$ in the expression for $\gamma_{\mathrm{ns}}^{(3)}(N)$ (green circles) are broken down to $\gamma_{\mathrm{rat}}^{(3)}(N)$ (purple rhombs), $\zeta_3\gamma_{\zeta_3}^{(3)}(N)$ (solid red triangles), $\zeta_4\gamma_{\zeta_4}^{(3)}(N)$ (yellow triangles), and $\zeta_5\gamma_{\zeta_5}^{(3)}(N)$ (blue squares) in QCD.}
\label{fig:nf01}
\end{center}
\end{figure}

In Fig.~\ref{fig:nf01}(a), the $n_f$-independent term in the expression for $\gamma_{\mathrm{ns}}^{(3)}(N)$ is broken down to $\gamma_{\mathrm{rat}}^{(3)}(N)$, $\zeta_3\gamma_{\zeta_3}^{(3)}(N)$, $\zeta_4\gamma_{\zeta_4}^{(3)}(N)$, and $\zeta_5\gamma_{\zeta_5}^{(3)}(N)$.
Looking at the transcendental contributions, we observe that the $\zeta_3$ contribution is positive, competes with the negative $\zeta_5$ contribution at low $N$ values, but wins out as $N$ increases, while the $\zeta_4$ contribution is throughout negligibly small.
On the other hand, the rational contribution is always dominant.

In Fig.~\ref{fig:nf01}(b), the consideration of Fig.~\ref{fig:nf01}(a) is repeated for the coefficient of $n_f$ in the expression for $\gamma_{\mathrm{ns}}^{(3)}(N)$.
Here, the $\zeta_3$ contribution is negative and always larger in magnitude than the positive $\zeta_4$ and $\zeta_5$ contributions.
The latter two almost compensate the negative rational contribution, so that the $\zeta_3$ contribution practically makes up the full result.

\begin{figure}
\begin{center}
\begin{tabular}{cc}
  \includegraphics[width=0.47\textwidth]{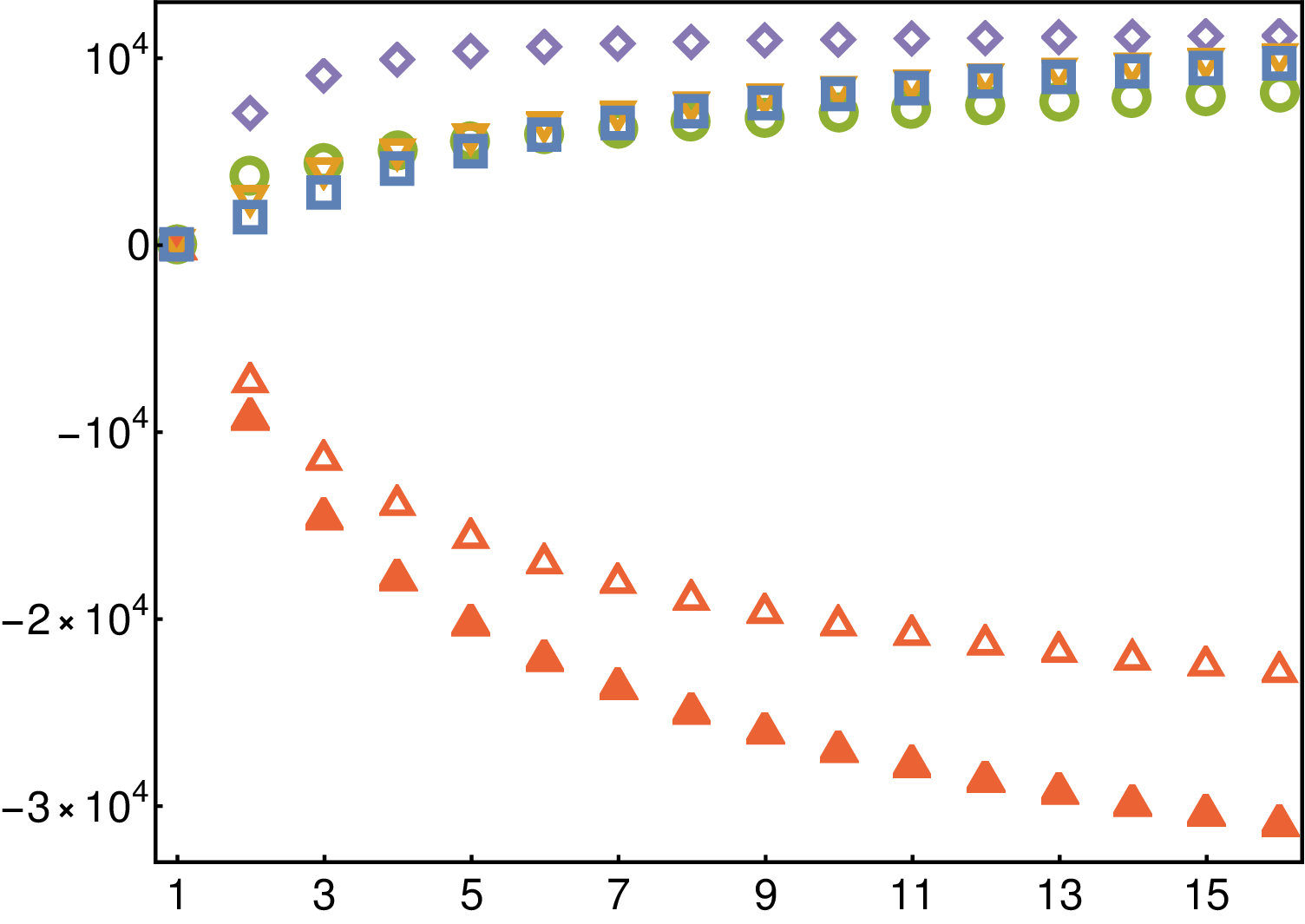}
  &
  \includegraphics[width=0.47\textwidth]{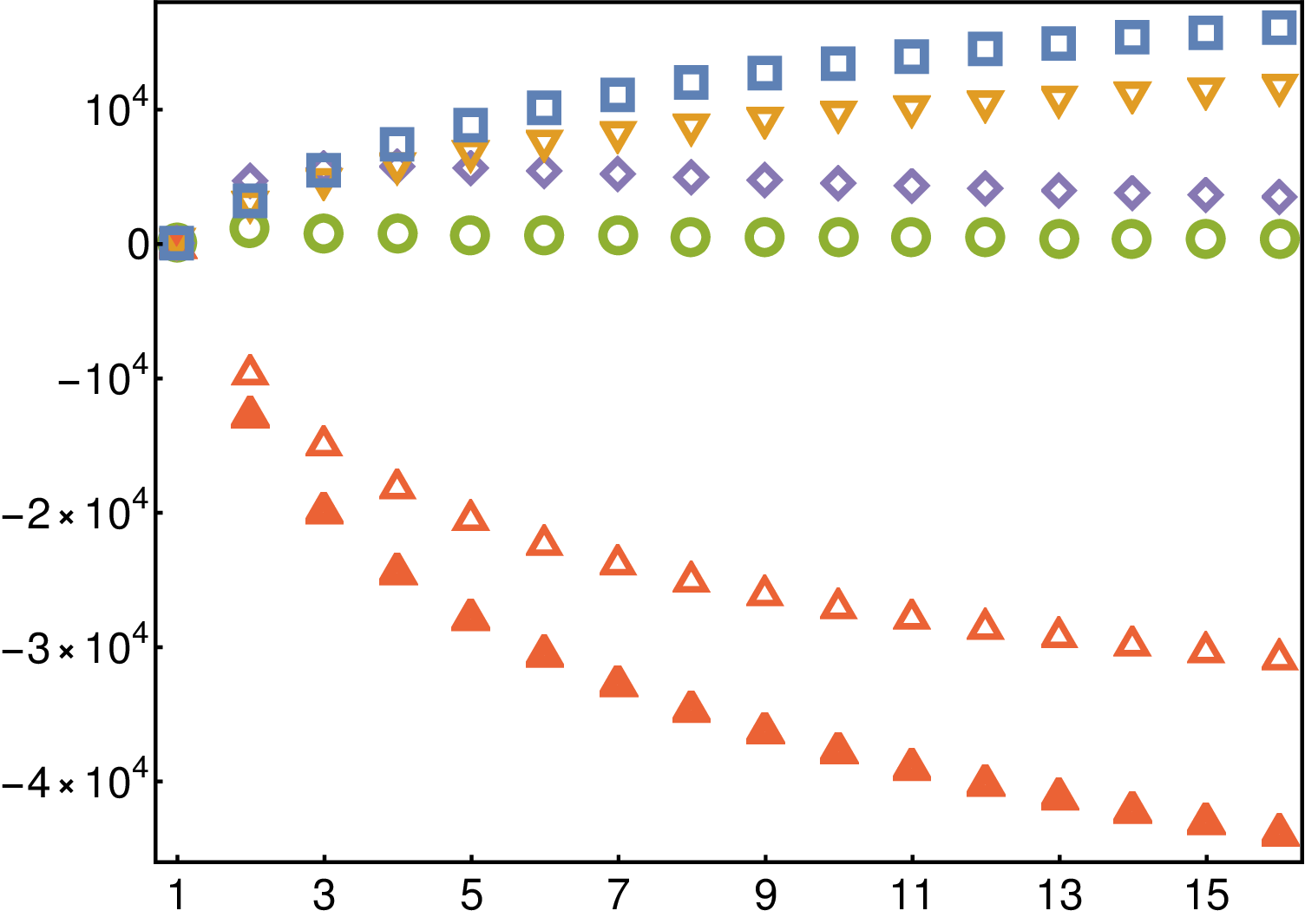}\\
(a) & (b)
\end{tabular}  
\caption{$\gamma_{\mathrm{ns}}^{(3)}(N)$ (green circles) is broken down to $\gamma_{\mathrm{rat}ry}^{(3)}(N)$ (purple rhombs), $\zeta_3\gamma_{\zeta_3}^{(3)}(N)$ (red triangles), $\zeta_4\gamma_{\zeta_4}^{(3)}(N)$ (yellow triangles), and $\zeta_5\gamma_{\zeta_5}^{(3)}(N)$ (blue squares) in QCD with (a) $n_f=4$ and (b) $n_f=5$.
The new part of $\zeta_3\gamma_{\zeta_3}^{(3)}(N)$ (solid red triangles), without the $n_f^3$ \cite{Gracey:1994nn} and $n_f^2$ \cite{Davies:2016jie} contributions, is shown for comparison.}
\label{fig:nf45}
\end{center}
\end{figure}

In Fig.~\ref{fig:nf45}(a) and (b), $\gamma_{\mathrm{ns}}^{(3)}(N)$ is broken down to $\gamma_{\mathrm{rat}}^{(3)}(N)$, $\zeta_3\gamma_{\zeta_3}^{(3)}(N)$, $\zeta_4\gamma_{\zeta_4}^{(3)}(N)$, and $\zeta_5\gamma_{\zeta_5}^{(3)}(N)$ in QCD with $n_f=4,5$, appropriate for EIC and LHC energies, respectively.
We observe that $\zeta_3\gamma_{\zeta_3}^{(3)}(N)$ is negative and larger in magnitude than any of the other terms, which are all positive.
Their superposition is highly destructive, yielding a small positive total for $n_f=4$ and approximately zero for $n_f=5$.
For comparison, the new contribution to $\zeta_3\gamma_{\zeta_3}^{(3)}(N)$, comprising the terms independent of and linear in $n_f$, is also shown.
It is seen to be enhanced with respect to $\zeta_3\gamma_{\zeta_3}^{(3)}(N)$, by about one third.

We thus conclude that our new results for $\gamma_{\zeta_3}^{(3)}(N)$ and $P_{\zeta_3}^{(3)\pm}(x)$ in QCD are numerically important.

\section{Conclusions}
\label{sec:con}

In our recent paper~\cite{Kniehl:2025jfs}, we considered the quark flavor non-singlet twist-two quark operator of arbitrary Lorentz spin $N$ in Eq.~\eqref{eq:nonsin} and extracted an all-$N$ expression for the term proportional to $\zeta(3)$ of its anomalous dimension $\gamma_{\mathrm{ns}}^{(3)\pm}(N)$ at four loops in SU($n_c$) color gauge theory from the published values for $N=1,\ldots,16$ \cite{Velizhanin:2011es,Velizhanin:2014fua,Moch:2017uml} via analytic reconstruction.
Specifically, we applied the LLL algorithm \cite{Lenstra82factoringpolynomials} as implemented in the program package \texttt{fplll} \cite{fplll}.
For lack of space, we listed in Ref.~\cite{Kniehl:2025jfs} only the QCD result, with $n_c=3$, and considered the respective pieces of the DGLAP splitting function $P_{\mathrm{ns}}^{(3)\pm}$ only in the limiting cases $x\to0,1$.
In the present paper, we reviewed our calculation and presented full results for general $n_c$, so as to expose their color structures.

Besides the quark flavor asymmetry, we also considered the valence configuration, extracted the transcendental parts, proportional to $\zeta(3)$ and $\zeta(5)$, of the respective four-loop anomalous dimension $\gamma_{\mathrm{ns}}^{(3)\mathrm{v}}(N)$ from its values for $N=3,5,\ldots,15$ \cite{Moch:2017uml}, and provided the corresponding pieces of the splitting function $P_{\mathrm{ns}}^{(3)\mathrm{v}}(x)$.

For the reader's convenience, all analytic results presented here are provided in machine-readable form in an ancillary file submitted along with this manuscript.

\section*{Note added}

After the submission of this paper, a preprint \cite{Gehrmann:2026qbl} appeared which provides complete analytic results for $\gamma_{\mathrm{ns}}^{(3)\pm}(N)$ and $\gamma_{\mathrm{ns}}^{(3)\mathrm{s}}(N)$ in ancillary files and confirms the $\zeta_5$ term in Eq.~\eqref{eq:vz5} and the $\zeta_3$ terms in Eqs.~\eqref{eq:vz3} and \eqref{eq:adN}.
Our conjecture regarding $\gamma_{\mathrm{rat}}^{(3)\mathrm{s}}(N)$ formulated below Eq.~\eqref{eq:fN} is validated by Ref.~\cite{Gehrmann:2026qbl} only as far as the first term of the combination $(1-8\eta+12\eta^2)$ is concerned.
Furthermore, the leading large-$N$ term of $\gamma_{\zeta_3}^{(3)\pm}(N)$, proportional to $\ln^2N$, is canceled by $\gamma_{\mathrm{rat}}^{(3)\pm}(N)$ \cite{Gehrmann:2026qbl} as expected.

\section*{Acknowledgments}

We thank S.-O. Moch and A. Vogt for comparing our all-$N$ results for $\gamma_{\zeta_3}^{(3)}(N)$, $\gamma_{\zeta_5}^{(3)\mathrm{s}}(N)$, and $\gamma_{\zeta_3}^{(3)\mathrm{s}}(N)$ in SU($n_c$) color gauge theory with theirs \cite{Moch:2025pri}.
This work was supported by the German Research Foundation DFG through Grant Nos.~KN~365/13-2 and 365/16-1.

\appendix

\boldmath
\section{Appendix: SU($n_c$) results in $N$ space}
\label{app:N}
\unboldmath

Here, we present the $\z3$ contribution $\gamma_{\z3}^{(3)}(N)$ to the anomalous dimension of the flavor non-singlet twist-two quark operator of Eq.~\eqref{eq:nonsin} for arbitrary Lorentz spin $N$ at four loops in SU($n_c$) color gauge theory with $n_f$ massless quark flavors, with the understanding that $\gamma_{\z3}^{(3)}(N)$ contributes to $\gamma_{\mathrm{ns}}^{(3)\pm}(N)$ for even/odd values of $N$.
We have
\footnotesize
\begin{eqnarray}
\gamma_{\z3}^{(3)}&=&
768\cf^4
\Big[
-\frac{35}{16}
-\frac{1}{36}(40-27\mm)\eta
+\frac{5}{8}(1-2\mm)\D1^2
-\frac{1}{24}(29-60\mm)\eta^2
+\frac{1}{12}(40+21\mm)\eta\HS_{1}
\nonumber\\&&{}
+\frac{1}{18}(9+\mm)\HS_{-2}
+\frac{5}{8}\HS_{2}
-\frac{1}{2}\mm\D1^3
-\frac{2}{9}(17-32\mm)\eta^3
-\mm\D1^2\HS_{1}
-\frac{1}{12}(25+153\mm)\eta^2\HS_{1}
\nonumber\\&&{}
+\frac{1}{18}(18+73\mm)\eta\HS_{-2}
-\frac{1}{6}(2+33\mm)\eta\HS_{2}
+\frac{2}{3}(1+15\mm)\eta\HS_{1,1}
-\frac{1}{4}(3-4\mm)\HS_{-3}
\nonumber\\&&{}
+\frac{1}{4}(3-2\mm)\HS_{3}
-\frac{1}{2}(3+4\mm)\HS_{-2,1}
-\frac{3}{2}\mm\D1^4
-\frac{1}{6}(9-16\mm)\eta^4
-2\mm\D1^3\HS_{1}
\nonumber\\&&{}
-\frac{1}{6}(1+37\mm)\eta^3\HS_{1}
+\D1^2\HS_{-2}
-\frac{1}{6}(9-20\mm)\eta^2\HS_{-2}
-\frac{1}{6}(1+33\mm)\eta^2\HS_{2}
\nonumber\\&&{}
+\frac{1}{3}(1+30\mm)\eta^2\HS_{1,1}
-\frac{1}{2}(9-2\mm)\eta\HS_{-3}
+\frac{1}{2}(2-\mm)\eta\HS_{3}
+(7-2\mm)\eta\HS_{-2,1}
\nonumber\\&&{}
-\frac{26}{3}\HS_{-4}
+\frac{10}{3}\HS_{4}
+19\HS_{-3,1}
-\frac{20}{3}\HS_{-2,-2}
+11\HS_{-2,2}
+9\HS_{1,-3}
-2\HS_{1,3}
+\HS_{2,-2}
\nonumber\\&&{}
+2\HS_{3,1}
-20\HS_{-2,1,1}
-14\HS_{1,-2,1}
\Big]
\nonumber\\&&{}
+256\cf^3\ca
\Big[
\frac{739}{96}
+\frac{1}{48}(533+405\mm)\eta
-\frac{1}{16}(45-46\mm)\D1^2
+\frac{1}{96}(907+156\mm)\eta^2
\nonumber\\&&{}
-\frac{1}{8}(186+167\mm)\eta\HS_{1}
-\frac{1}{6}(105+2\mm)\HS_{-2}
-\frac{45}{16}\HS_{2}
-\frac{13}{4}\mm\D1^3
+\frac{5}{24}(112-183\mm)\eta^3
\nonumber\\&&{}
+\frac{9}{2}\mm\D1^2\HS_{1}
+\frac{1}{8}(161+535\mm)\eta^2\HS_{1}
-\frac{7}{6}(9+26\mm)\eta\HS_{-2}
+\frac{1}{4}(24+131\mm)\eta\HS_{2}
\nonumber\\&&{}
-(12+61\mm)\eta\HS_{1,1}
-\frac{3}{8}(11+20\mm)\HS_{-3}
-\frac{1}{24}(23-90\mm)\HS_{3}
+\frac{1}{4}(131+60\mm)\HS_{-2,1}
\nonumber\\&&{}
+\frac{27}{4}\mm\D1^4
+\frac{1}{8}(85-124\mm)\eta^4
+9\mm\D1^3\HS_{1}
+\frac{5}{4}(2+29\mm)\eta^3\HS_{1}
-\frac{9}{2}\D1^2\HS_{-2}
\nonumber\\&&{}
+\frac{1}{4}(27-104\mm)\eta^2\HS_{-2}
+\frac{1}{4}(12+131\mm)\eta^2\HS_{2}
-(6+61\mm)\eta^2\HS_{1,1}
+\frac{5}{4}(17-6\mm)\eta\HS_{-3}
\nonumber\\&&{}
-\frac{5}{4}(4-3\mm)\eta\HS_{3}
-\frac{1}{2}(67-30\mm)\eta\HS_{-2,1}
+45\HS_{-4}
-24\HS_{4}
-\frac{207}{2}\HS_{-3,1}
+52\HS_{-2,-2}
\nonumber\\&&{}
-\frac{131}{2}\HS_{-2,2}
-\frac{85}{2}\HS_{1,-3}
+10\HS_{1,3}
-\frac{9}{2}\HS_{2,-2}
-14\HS_{3,1}
+122\HS_{-2,1,1}
+67\HS_{1,-2,1}
\Big]
\nonumber\\&&{}
+256\cf^2\ca^2
\Big[
\frac{137}{128}
-\frac{1}{32}(398+497\mm)\eta
+\frac{3}{16}(5+12\mm)\D1^2
+\frac{1}{24}(394+455\mm)\eta\HS_{1}
\nonumber\\&&{}
-\frac{1}{48}(274+713\mm)\eta^2
+\frac{1}{4}(86+\mm)\HS_{-2}
+\frac{15}{16}\HS_{2}
+\frac{15}{2}\D1^3\mm
-\frac{1}{48}(697-994\mm)\eta^3
\nonumber\\&&{}
-\frac{3}{2}\D1^2\HS_{1}\mm
+\frac{1}{12}(103+297\mm)\eta\HS_{-2}
-\frac{1}{4}(34+89\mm)\eta\HS_{2}
+(17+43\mm)\eta\HS_{1,1}
\nonumber\\&&{}
+\frac{1}{24}(109+144\mm)\HS_{-3}
-\frac{1}{24}(73+72\mm)\HS_{3}
-\frac{1}{12}(419+144\mm)\HS_{-2,1}
+\frac{29}{6}\HS_{1,-2}
\nonumber\\&&{}
-\frac{9}{4}\D1^4\mm
-\frac{1}{2}(16-19\mm)\eta^4
-3\D1^3\HS_{1}\mm
+\frac{3}{2}\D1^2\HS_{-2}
-\frac{1}{2}(8-43\mm)\eta^2\HS_{-2}
\nonumber\\&&{}
-\frac{1}{24}(502+883\mm)\eta^2\HS_{1}
-\frac{1}{4}(17+89\mm)\eta^2\HS_{2}
+\frac{1}{2}(17+86\mm)\eta^2\HS_{1,1}
-\frac{1}{4}(13+89\mm)\eta^3\HS_{1}
\nonumber\\&&{}
-\frac{1}{4}(43-24\mm)\eta\HS_{-3}
+\frac{1}{2}(5-6\mm)\eta\HS_{3}
+\frac{3}{2}(11-8\mm)\eta\HS_{-2,1}
+2\eta\HS_{1,-2}
-25\HS_{-4}
\nonumber\\&&{}
+19\HS_{4}
+\frac{125}{2}\HS_{-3,1}
+10\HS_{3,1}
-43\HS_{-2,-2}
+\frac{89}{2}\HS_{-2,2}
+\frac{43}{2}\HS_{1,-3}
-5\HS_{1,3}
+\frac{7}{2}\HS_{2,-2}
\nonumber\\&&{}
-86\HS_{-2,1,1}
-33\HS_{1,-2,1}
-4\HS_{1,1,-2}
\Big]
\nonumber\\&&{}
+\frac{64}{3}\cf\ca^3
\Big[
-\frac{1351}{48}
+\frac{1}{72}(4384+4311\mm)\eta
+\frac{419}{36}\HS_{1}
-\frac{33}{2}\D1^2\mm
+\frac{1}{12}(107+865\mm)\eta^2
\nonumber\\&&{}
-\frac{1}{4}(236+263\mm)\eta\HS_{1}
-\frac{5}{12}(165+2\mm)\HS_{-2}
-33\D1^3\mm
+\frac{1}{4}(128-161\mm)\eta^3
\nonumber\\&&{}
-\frac{1}{12}(247+946\mm)\eta\HS_{-2}
+(23+64\mm)\eta\HS_{2}
-2(23+64\mm)\eta\HS_{1,1}
+\frac{1}{4}(317+312\mm)\eta^2\HS_{1}
\nonumber\\&&{}
-\frac{1}{6}(86+117\mm)\HS_{-3}
+\frac{1}{12}(235+117\mm)\HS_{3}
+\frac{1}{2}(263+78\mm)\HS_{-2,1}
-\frac{221}{6}\HS_{1,-2}
\nonumber\\&&{}
+\frac{45}{2}(1-\mm)\eta^4
+\frac{1}{4}(74+203\mm)\eta^3\HS_{1}
+(13-68\mm)\eta^2\HS_{-2}
+\frac{1}{2}(47+128\mm)\eta^2\HS_{2}
\nonumber\\&&{}
-(47+128\mm)\eta^2\HS_{1,1}
+\frac{1}{2}(37-39\mm)\eta\HS_{-3}
-\frac{1}{4}(2-39\mm)\eta\HS_{3}
-3(7-13\mm)\eta\HS_{-2,1}
\nonumber\\&&{}
-16\eta\HS_{1,-2}
+54\HS_{-4}
-54\HS_{4}-149\HS_{-3,1}
+136\HS_{-2,-2}
-128\HS_{-2,2}
-37\HS_{1,-3}
+\HS_{1,3}
\nonumber\\&&{}
-16\HS_{2,-2}
-29\HS_{3,1}
+256\HS_{-2,1,1}
+42\HS_{1,-2,1}
+32\HS_{1,1,-2}
\Big]
\nonumber\\&&{}
+512\frac{d_{44}^{\mathrm{RA}}}{\nr}
\Big[
\frac{55}{96}
-\frac{1}{8}(70-37\mm)\eta
+\frac{1}{12}\HS_{1}
-\frac{1}{4}(1-27\mm)\eta\HS_{1}
+\frac{1}{12}(34+11\mm)\eta^2
\nonumber\\&&{}
-\frac{1}{12}(120-\mm)\HS_{-2}
-\frac{1}{12}(47-37\mm)\eta\HS_{-2}
-\frac{1}{2}(7+8\mm)\eta\HS_{2}
+(7+8\mm)\eta\HS_{1,1}
\nonumber\\&&{}
-\frac{1}{2}(13+\mm)\eta^2\HS_{1}
+\frac{13}{8}(1-\mm)\eta^3
-\frac{1}{12}(61+9\mm)\HS_{3}
+\frac{1}{6}(17+9\mm)\HS_{-3}
-\frac{3}{2}(9+2\mm)\HS_{-2,1}
\nonumber\\&&{}
+\frac{47}{6}\HS_{1,-2}
+\frac{1}{2}(5+3\mm)\eta\HS_{-3}
-\frac{1}{4}(10+3\mm)\eta\HS_{3}
-3(3+\mm)\eta\HS_{-2,1}
+4\eta\HS_{1,-2}
\nonumber\\&&{}
-\frac{1}{2}(5-4\mm)\eta^2\HS_{-2}
-\frac{1}{4}(1+16\mm)\eta^2\HS_{2}
+\frac{1}{2}(1+16\mm)\eta^2\HS_{1,1}
+\frac{7}{4}(1+\mm)\eta^3\HS_{1}
-\HS_{-3,1}
\nonumber\\&&{}
-4\HS_{-2,-2}
+8\HS_{-2,2}
-5\HS_{1,-3}
+5\HS_{1,3}
+4\HS_{2,-2}
-\HS_{3,1}
-16\HS_{-2,1,1}
+18\HS_{1,-2,1}
-8\HS_{1,1,-2}
\Big]\nonumber\\&&{}
+\frac{64}{3}\cf^3\nf
\Big[
-\frac{53}{16}
-\frac{1}{8}(61+180\mm)\eta
+\frac{37}{4}\HS_{1}
+12\eta(1+\mm)\HS_{1}
+3(1+2\mm)\D1^2
+(1-27\mm)\eta^2
\nonumber\\&&{}
+30\HS_{-2}
+9\HS_{2}
-6(1-2\mm)\D1^3
+12\eta\HS_{-2}
+6\eta\HS_{2}
-12\D1^2\HS_{1}
+6(1+2\mm)\eta^2\HS_{1}
\nonumber\\&&{}
+(4-5\mm)\eta^3
+12\HS_{-3}
-2\HS_{3}
-24\HS_{-2,1}
-24\HS_{1,-2}
-12\HS_{1,2}
-12\HS_{2,1}
\Big]
\nonumber\\&&{}
+\frac{64}{9}\cf^2\ca\nf
\Big[
-\frac{705}{16}
-\frac{1}{8}(169-774\mm)\eta
+\frac{241}{4}\HS_{1}
-6(7+8\mm)\eta\HS_{1}
-\frac{3}{2}(17+18\mm)\D1^2
\nonumber\\&&{}
-\frac{3}{4}(21-152\mm)\eta^2
-126\HS_{-2}
-\frac{87}{2}\HS_{2}
-48\eta\HS_{-2}
-18\eta\HS_{2}
+36\D1^2\HS_{1}
-3(7+16\mm)\eta^2\HS_{1}
\nonumber\\&&{}
+18(1-3\mm)\D1^3
-\frac{3}{2}(16-13\mm)\eta^3
-42\HS_{-3}
+33\HS_{3}
+96\HS_{-2,1}
+96\HS_{1,-2}
+36\HS_{1,2}
+36\HS_{2,1}
\Big]
\nonumber\\&&{}
+\frac{64}{3}\cf\ca^2\nf
\Big[
\frac{1789}{96}
-\frac{385}{12}\HS_{1}
+\frac{1}{8}(113-83\mm)\eta
+5(1+\mm)\eta\HS_{1}
+\frac{1}{2}(11+6\mm)\D1^2
\nonumber\\&&{}
+\frac{1}{24}(74-293\mm)\eta^2
+\frac{53}{4}\HS_{-2}
+\frac{11}{2}\HS_{2}
+6\mm\D1^3
+\frac{1}{12}(41-23\mm)\eta^3
+\frac{29}{6}\eta\HS_{-2}
+2\eta\HS_{2}
\nonumber\\&&{}
-4\eta\HS_{1,1}
+5\mm\eta^2\HS_{1}
+\frac{23}{6}\HS_{-3}
-\frac{41}{6}\HS_{3}
-10\HS_{-2,1}
-\frac{29}{3}\HS_{1,-2}
-2\eta^3\HS_{1}
-2\eta^2\HS_{2}
+4\eta^2\HS_{1,1}
\Big]
\nonumber\\&&{}
+512\frac{d_{44}^{\mathrm{RR}}}{\nr}\nf
\Big[
\frac{53}{48}
-\frac{1}{12}(22+3\mm)\eta
-\frac{1}{6}\HS_{1}
+2\eta\HS_{1}
+\frac{1}{12}(10-\mm)\eta^2
+\frac{1}{2}\HS_{-2}
+\frac{1}{3}\eta\HS_{-2}
\nonumber\\&&{}
+2\eta\HS_{2}
-4\eta\HS_{1,1}
+\frac{1}{6}(1-\mm)\eta^3
-\frac{1}{3}\HS_{3}
+\frac{1}{3}\HS_{-3}
-\frac{2}{3}\HS_{1,-2}
-2\eta^2\HS_{2}
+4\eta^2\HS_{1,1}
-2\eta^3\HS_{1}
\Big]
\,,
\label{eq:adN}
\end{eqnarray}
\normalsize
where $\varepsilon$, $\eta$, and $D_1$ have been defined along with Eqs.~\eqref{eq:zeta3}, \eqref{AnsatzOddEven}, and \eqref{eq:Nqcd}, respectively, and the argument $N$ has been dropped in the NHSs.

For $N\to0+\w$, we have
\footnotesize
\begin{eqnarray}
\gamma_{\z3}^{(3)+}&=&
\frac{32}{\w^4} \cf^2 (28 \cf^2 - 39 \cf \ca + 12 \ca^2)
+\frac{16}{3\w^3} \cf \big(-192 \cf^3 + 226 \cf^2 \ca + 9 \cf \ca^2 - 33 \ca^3 - 4 \cf^2 \nf 
\nonumber\\&&
- 6 \cf \ca \nf + 6 \ca^2 \nf
\big)
+ \frac{8}{3\w^2}\Big[ 
12 \cf^4 (71 - 152 \z2) 
+ \cf^3 \ca (643 + 3720 \z2) 
- 12 \cf^2 \ca^2 (193 + 204 \z2) 
\nonumber\\&&
+ 2 \cf \ca^3 (423 + 277 \z2) 
+ 48 \frac{d_{44}^{\mathrm{RA}}}{\nr} (15 + 14 \z2) 
-184 \cf^3 \nf 
+ 298 \cf^2 \ca \nf 
- \cf \ca^2 \nf (109 + 16 \z2) 
\nonumber\\&&
+48 \frac{d_{44}^{\mathrm{RR}}}{\nr}\nf(3 - 8 \z2) 
\Big]
+ \frac{8}{27\w} \Big[
72 \cf^4 (-226+150 \z2  + 93 \z3) 
+ 18 \cf^3 \ca (295 - 1404 \z2 - 258 \z3) 
\nonumber\\&&
+ 9 \cf^2 \ca^2 (1947 + 1804 \z2 - 216 \z3) 
- \cf \ca^3 (6533 + 3636 \z2 - 954 \z3) 
- 216 \frac{d_{44}^{\mathrm{RA}}}{\nr}(93 + 140 \z2 + 22 \z3)  
\nonumber\\&&
+ 9 \cf^3 \nf (127 + 144 \z2) 
- 3 \cf^2 \ca \nf (1183 + 552 \z2) 
+ 72 \cf \ca^2 \nf (31 + 11 \z2 + 2 \z3) 
\nonumber\\&&
- 144 \frac{d_{44}^{\mathrm{RR}}}{\nr}\nf (43 - 72 \z2 - 24 \z3) 
\Big]\,,
\nonumber\\
\gamma_{\z3}^{(3)-}&=&
\frac{32}{\w^4}\cf \big(
-100 \cf^3 
+ 209 \cf^2 \ca 
- 140 \cf \ca^2 
+ 30 \ca^3
\big)
+ \frac{16}{9\w^3}\Big(
2496 \cf^4 
- 6198 \cf^3 \ca 
+ 5007 \cf^2 \ca^2 
\nonumber\\&&
- 1293 \cf \ca^3 
+ 936 \frac{d_{44}^{\mathrm{RA}}}{\nr} 
+ 108 \cf^3 \nf 
- 174 \cf^2 \ca \nf 
+ 64 \cf \ca^2 \nf 
+ 96  \frac{d_{44}^{\mathrm{RR}}}{\nr}\nf
\Big)
\nonumber\\&&
+ \frac{8}{9\w^2}\Big[
36 \cf^4 (-305+ 144 \z2) 
+ 3 \cf^3 \ca (8131 - 3240 \z2) 
- 912 \cf^2 \ca^2 (19 - 6 \z2) 
\nonumber\\&&
+ 2 \cf \ca^3 (2041 - 387 \z2) 
- 4512 \frac{d_{44}^{\mathrm{RA}}}{\nr}
+ 24 \cf^3 \nf 
+ 6 \cf^2 \ca \nf 
- \cf \ca^2 \nf (17 + 48 \z2) 
\nonumber\\&&
- 48 \frac{d_{44}^{\mathrm{RR}}}{\nr}\nf (1 + 24 \z2) 
\Big]
+ \frac{8}{27\w}\Big[
72 \cf^4 (848 - 264 \z2 + 81 \z3) 
- 18 \cf^3 \ca (8003 - 2616 \z2 + 954 \z3) 
\nonumber\\&&
+ 9 \cf^2 \ca^2 (11849 - 3948 \z2 + 1992 \z3) 
- \cf \ca^3 (24419 - 7056 \z2 + 6426 \z3) 
\nonumber\\&&
+ 72 \frac{d_{44}^{\mathrm{RA}}}{\nr} (55 - 144 \z2 + 162 \z3) 
+ 9 \cf^3 \nf (103 - 48 \z2) 
- 3 \cf^2 \ca \nf (955 - 216 \z2) 
\nonumber\\&&
+ 6 \cf \ca^2 \nf (311 + 12 \z2 + 24 \z3) 
- 144 \frac{d_{44}^{\mathrm{RR}}}{\nr}\nf (17 - 72 \z2 - 24 \z3) 
\Big]\,.
\end{eqnarray}
\normalsize

For $N\to\infty$, we have
\footnotesize
\begin{eqnarray}
\gamma_{\z3}^{(3)\pm}&=&
\frac{256}{3} \z2 \Big(\ln^2N +\frct{\ln N}{N}\Big)\Big(
 3  \cf^2\ca^2 
-2 \cf\ca^3  
+ 12  \frac{d_{44}^{\mathrm{RA}}}{\nr}
\Big)
+\frac{16}{27}\ln N \Big[
- 1044\z2  \cf^2\ca^2  
\nonumber\\&&
+  \cf\ca^3 (419 + 663 \z2 - 54 \z3) 
+ 72 \frac{d_{44}^{\mathrm{RA}}}{\nr}(1 - 47 \z2 - 18 \z3)  
+333 \cf^3 \nf 
\nonumber\\&&
+ 3 \cf^2\ca  \nf (241 - 48 \z2) 
- 3  \cf\ca^2 \nf (385 - 58 \z2) 
- 144 \frac{d_{44}^{\mathrm{RR}}}{\nr}\nf(1 - 2 \z2) 
\Big]
\nonumber\\&&
+ \frac{2}{9} \Big\{
- 24 \cf^4 [315 - 2(27 - 2 \mm) \z2 - 324 \z3 + 108 \z4] 
+ 12 \cf^3\ca  [739 + 2(285 + 8 \mm) \z2
\nonumber\\&&
  - 1760 \z3 + 432 \z4] 
+ 3  \cf^2\ca^2 [411 - 24(157+2 \mm) \z2 + 5672 \z3 - 1080 \z4] 
\nonumber\\&&
- 2  \cf\ca^3 [1351 - 10(165+2 \mm) \z2 + 2268 \z3 - 492 \z4] 
+ 24 \frac{d_{44}^{\mathrm{RA}}}{\nr}[55 + 4(120 - \mm) \z2 + 24 \z3
\nonumber\\&&
  - 336 \z4]  
- 6 \cf^3 \nf (53 + 96 \z2 + 80 \z3) 
- 6 \cf^2\ca  \nf (235 - 104 \z2 - 152 \z3) 
\nonumber\\&&
+ \cf\ca^2 \nf (1789 - 108 \z2 - 216 \z3) 
+ 48 \frac{d_{44}^{\mathrm{RR}}}{\nr}\nf(53 - 12 \z2 - 24 \z3)  
\Big\} 
\nonumber\\&&
+\frac{8}{27N} \Big[
- 1908\z2  \cf^2\ca^2 
+ \ca^3 \cf (419 + 1239 \z2 - 54 \z3) 
+ 72 \frac{d_{44}^{\mathrm{RA}}}{\nr}(1 - 95 \z2 - 18 \z3)   
\nonumber\\&&
  +333 \cf^3 \nf 
+ 3\cf^2 \ca  \nf (241 - 48 \z2) 
- 3  \cf\ca^2 \nf (385 - 58 \z2) 
- 144 \frac{d_{44}^{\mathrm{RR}}}{\nr}\nf(1 - 2 \z2)  
\Big]\,.
\nonumber\\&&
\label{eq:largeN}
\end{eqnarray}
\normalsize
The leading large-$N$ term proportional to $\ln^2N$ in Eq.~\eqref{eq:largeN} is in conflict with the bound on $\gamma_{\mathrm{ns}}^{(n)\pm}(N)$ established in Ref.~\cite{Korchemsky:1988si} and must cancel against similar terms from other components of Eq.~\eqref{eq:zeta3}, including the one from $\gamma_{\zeta_5}^{(3)}(N)$ given in Eq.~(7.1) of Ref.~\cite{Moch:2017uml}.
However, a structure across $\gamma_{\zeta_5}^{(3)}(N)$, $\gamma_{\zeta_3}^{(3)}(N)$, and $\gamma_{\mathrm{rat}}^{(3)}(N)$ via $f(N)$ of Eq.~\eqref{eq:fN}, similar to the case of $\gamma_{\mathrm{ns}}^{(3)\mathrm{s}}(N)$, albeit compatible with Eq.~\eqref{eq:adN}, is incompatible with what is known about $\gamma_{\mathrm{rat}}^{(3)}(N)$ from the planar \cite{Beisert:2006ez,Bajnok:2008qj} and nonplanar \cite{Kniehl:2020rip,Kniehl:2021ysp,Kniehl:2024tvd} universal anomalous dimensions of the twist-two operators with general Lorentz spin $N$ at NNNLO in $\mathcal{N}=4$ SYM by the maximal-transcendentality principle \cite{Kotikov:2002ab,Kotikov:2006ts}.

\boldmath
\section{Appendix: SU($n_c$) results in $x$ space}
\label{app:x}
\unboldmath

Here, we present the parts of splitting function that emerge from Eq.~\eqref{eq:adN}, for either even or odd values of $N$, via Mellin transformation as described in Section~\ref{sec:x}.
They read
\footnotesize
\begin{eqnarray}
P_{\z3}^{(3)+}&=&
128\cf^4
\Big[
\frac{113}{3}(1-x)
-\frac{1}{6}(87-259x)\z2
+16\z3
+\frac{1}{24}(511-137x)\HPL_{0}
+(25-9x)\z2\HPL_{0}
\nonumber\\&&{}
+\frac{39}{2}(1-x)\HPL_{1}
-12(1+x)\z2\HPL_{-1}
+\frac{1}{6}(39-64x)\HPL_{0,0}
+\frac{1}{2}(29-103x)\HPL_{0,1}
-\frac{25}{3}(1+x)\HPL_{-1,0}
\nonumber\\&&{}
+33(1-x)\HPL_{1,0}
+60(1-x)\HPL_{1,1}
+6(1+x)\HPL_{-1,0,0}
+12(1+x)\HPL_{-1,0,1}
-6(1-x)\HPL_{0,-1,0}
\nonumber\\&&{}
-(9-11x)\HPL_{0,0,0}
-5(5-3x)\HPL_{0,0,1}
+(1+x)\HPL_{0,1,0}
+2(1+x)\HPL_{0,1,1}
+3(1-x)\HPL_{1,0,0}
\nonumber\\&&{}
+\frac{3}{2}\pqqp\Big(
20\z3
-\frac{5}{4}\HPL_{0}
+\z2\frac{11}{3}\HPL_{0}
+\frac{1}{2}\HPL_{0,0}
+\frac{40}{3}\HPL_{0,-1,0}
-\frac{20}{3}\HPL_{0,0,0}
+4\HPL_{0,0,1}
-4\HPL_{1,0,0}
\Big)
\nonumber\\&&{}
+\pqqm\Big(
-\frac{21}{2}\z2
-\frac{123}{2}\z3
-\frac{137}{2}\z2\HPL_{0}
+42\z2\HPL_{-1}
-27\HPL_{-1,0,0}
-42\HPL_{-1,0,1}
-3\HPL_{0,-1,0}
\nonumber\\&&{}
+26\HPL_{0,0,0}
+57\HPL_{0,0,1}
+33\HPL_{0,1,0}
+60\HPL_{0,1,1}
+\frac{3}{4}\HPL_{0,0}
-\frac{5}{3}\HPL_{0}
+\frac{21}{2}\HPL_{0,1}
\Big)
+\frac{1}{24}(315-50\z2
\nonumber\\&&{}
-324\z3+108\z4)\delta(1-x)
\Big]
\nonumber\\&&{}
 +256\cf^3\ca
\Big[
-\frac{295}{48}(1-x)
+\frac{1}{24}(129-1387x)\z2
-\frac{1}{4}(53+13x)\z3
-\frac{1}{8}(29-109x)\HPL_{0}
\nonumber\\&&{}
-\frac{1}{2}(40-13x)\z2\HPL_{0}
-\frac{1}{24}(7-500x)\HPL_{0,0}
-\frac{1}{8}(43-481x)\HPL_{0,1}
+\frac{7}{3}(1+x)\HPL_{-1,0}
-\frac{115}{8}(1-x)\HPL_{1}
\nonumber\\&&{}
+15(1+x)\z2\HPL_{-1}
-\frac{131}{4}(1-x)\HPL_{1,0}
-61(1-x)\HPL_{1,1}
+\frac{1}{8}(45-67x)\HPL_{0,0,0}
+\frac{9}{2}(1-x)\HPL_{0,-1,0}
\nonumber\\&&{}
-\frac{15}{2}(1+x)\HPL_{-1,0,0}
-\frac{15}{4}(1-x)\HPL_{1,0,0}
-15(1+x)\HPL_{-1,0,1}
+(20-11x)\HPL_{0,0,1}
-3(1+x)\HPL_{0,1,0}
\nonumber\\&&{}
-6(1+x)\HPL_{0,1,1}
-\frac{1}{4}\pqqm
\Big(
-\frac{191}{2}\z2
-\frac{445}{2}\z3
-\frac{107}{3}\HPL_{0}
-\frac{527}{2}\z2\HPL_{0}
+134\z2\HPL_{-1}
+\frac{93}{4}\HPL_{0,0}
\nonumber\\&&{}
+\frac{191}{2}\HPL_{0,1}
-85\HPL_{-1,0,0}
-134\HPL_{-1,0,1}
-9\HPL_{0,-1,0}
+90\HPL_{0,0,0}
+207\HPL_{0,0,1}
+131\HPL_{0,1,0}
+244\HPL_{0,1,1}
\Big)
\nonumber\\&&{}
-\pqqp\Big(
37\z3
-\frac{45}{32}\HPL_{0}
+\frac{57}{8}\z2\HPL_{0}
-\frac{67}{48}\HPL_{0,0}
+26\HPL_{0,-1,0}
-12\HPL_{0,0,0}
+7\HPL_{0,0,1}
-5\HPL_{1,0,0}
\Big)
\nonumber\\&&{}
-\frac{1}{96}(739+586\z2-1760\z3+432\z4)\delta(1-x)
\Big]
\nonumber\\&&{}
+64\cf^2\ca^2
\Big[
-\frac{649}{8}(1-x)
+\frac{7}{3}(8+83x)\z2
+\frac{1}{2}(59+17x)\z3
\nonumber\\&&{}
-\frac{1}{24}(1227+1223x)\HPL_{0}
+\frac{53}{6}(1-x)\HPL_{1}
+(43-20x)\z2\HPL_{0}
-48(1+x)\z2\HPL_{-1}
-3(1+x)\HPL_{-1,0}
\nonumber\\&&{}
-\frac{1}{12}(117+1067x)\HPL_{0,0}
+89(1-x)\HPL_{1,0}
-\frac{2}{3}(28+295x)\HPL_{0,1}
+172(1-x)\HPL_{1,1}
-6(1-3x)\HPL_{0,0,0}
\nonumber\\&&{}
+24(1+x)\HPL_{-1,0,0}
+48(1+x)\HPL_{-1,0,1}
-9(1-x)\HPL_{0,-1,0}
-(43-29x)\HPL_{0,0,1}
+17(1+x)\HPL_{0,1,0}
\nonumber\\&&{}
+34(1+x)\HPL_{0,1,1}
+12(1-x)\HPL_{1,0,0}
+\pqqp\Big(
-\frac{29}{6}\z2
+119\z3
-\frac{15}{8}\HPL_{0}
+4\z2\HPL_{1}
+\frac{53}{2}\z2\HPL_{0}
\nonumber\\&&{}
-\frac{145}{12}\HPL_{0,0}
+86\HPL_{0,-1,0}
-38\HPL_{0,0,0}
+20\HPL_{0,0,1}
-10\HPL_{1,0,0}
\Big)
+\pqqm
\Big(
-89\z2
-\frac{279}{2}\z3
-\frac{343}{2}\z2\HPL_{0}
\nonumber\\&&{}
-\frac{87}{2}\HPL_{0}
+70\z2\HPL_{-1}
+\frac{253}{12}\HPL_{0,0}
+\frac{563}{6}\HPL_{0,1}
+\frac{29}{3}\HPL_{-1,0}
+8\HPL_{-1,-1,0}
-43\HPL_{-1,0,0}
-66\HPL_{-1,0,1}
-7\HPL_{0,-1,0}
\nonumber\\&&{}
+50\HPL_{0,0,0}
+125\HPL_{0,0,1}
+89\HPL_{0,1,0}
+172\HPL_{0,1,1}
\Big)
-\frac{1}{96}(411-3816\z2+5672\z3-2040\z4)\delta(1-x)
\Big]
\nonumber\\&&{}
+\frac{32}{3}\cf\ca^3
\Big[
\frac{1019}{6}(1-x)
-\frac{1}{6}(417+2435x)\z2
-\frac{1}{2}(49-3x)\z3
+\frac{1}{12}(1703+737x)\HPL_{0}
\nonumber\\&&{}
-\frac{1}{2}(79-91x)\z2\HPL_{0}
+\frac{95}{2}(1-x)\HPL_{1}
+78(1+x)\z2\HPL_{-1}
+\frac{7}{6}(18+179x)\HPL_{0,0}
-176(1-x)\HPL_{1,0}
\nonumber\\&&{}
+\frac{1}{2}(139+843x)\HPL_{0,1}
-352(1-x)\HPL_{1,1}
-39(1+x)\HPL_{-1,0,0}
-78(1+x)\HPL_{-1,0,1}
-18x\HPL_{0,0,0}
\nonumber\\&&{}
+10(1-x)\HPL_{0,-1,0}
-47(1+x)\HPL_{0,1,0}
-94(1+x)\HPL_{0,1,1}
+\frac{1}{2}(79-111x)\HPL_{0,0,1}
-\frac{39}{2}(1-x)\HPL_{1,0,0}
\nonumber\\&&{}
+\frac{47}{3}(1+x)\HPL_{-1,0}
-\pqqp
\Big(
-\frac{419}{36}
-\frac{221}{12}\z2
+\frac{355}{2}\z3
-\frac{88}{3}\HPL_{0,0}
+47\z2\HPL_{0}
+16\z2\HPL_{1}
\nonumber\\&&{}
+136\HPL_{0,-1,0}
-54\HPL_{0,0,0}
+29\HPL_{0,0,1}
-\HPL_{1,0,0}
\Big)
-8\pqqm
\Big(
-\frac{1825}{96}\z2
-\frac{351}{16}\z3
-\frac{835}{96}\HPL_{0}
\nonumber\\&&{}
-\frac{225}{8}\z2\HPL_{0}
+\frac{29}{4}\z2\HPL_{-1}
+\frac{203}{48}\HPL_{0,0}
+\frac{341}{16}\HPL_{0,1}
+\frac{221}{48}\HPL_{-1,0}
+4\HPL_{-1,-1,0}
-2\HPL_{0,-1,0}
+16\HPL_{0,1,0}
\nonumber\\&&{}
+32\HPL_{0,1,1}
-\frac{21}{4}\HPL_{-1,0,1}
+\frac{27}{4}\HPL_{0,0,0}
-\frac{37}{8}\HPL_{-1,0,0}
+\frac{149}{8}\HPL_{0,0,1}
\Big)
\nonumber\\&&{}
+\frac{1}{24}(1351-1670\z2+2268\z3-1452\z4)\delta(1-x)
\Big]
\nonumber\\&&{}
+128\frac{d_{44}^{\mathrm{RA}}}{\nr}
\Big[
\frac{139}{3}(1-x)
+\frac{1}{3}(309+284x)\z2
+(7+9x)\z3
\nonumber\\&&{}
-\frac{1}{6}(29-967x)\HPL_{0}
+(10+2x)\z2\HPL_{0}
+\frac{1}{3}(9-281x)\HPL_{0,0}
-12(1+x)\z2\HPL_{-1}
-166(1-x)\HPL_{1}
\nonumber\\&&{}
-(103+111x)\HPL_{0,1}
+4(1-x)\HPL_{1,0}
+8(1-x)\HPL_{1,1}
+6(1+x)\HPL_{-1,0,0}
+12(1+x)\HPL_{-1,0,1}
\nonumber\\&&{}
-2(1-x)\HPL_{0,-1,0}
-10\HPL_{0,0,1}
+(1+x)\HPL_{0,1,0}
+2(1+x)\HPL_{0,1,1}
+3(1-x)\HPL_{1,0,0}
-\frac{49}{3}(1+x)\HPL_{-1,0}
\nonumber\\&&{}
+2\pqqp\Big(
\frac{1}{12}
-\frac{47}{12}\z2
+\frac{1}{2}\z3
+5\z2\HPL_{0}
+4\z2\HPL_{1}
-\frac{35}{6}\HPL_{0,0}
+4\HPL_{0,-1,0}
-\HPL_{0,0,1}
+5\HPL_{1,0,0}
\Big)
\nonumber\\&&{}
-8\pqqm\Big(
\frac{151}{48}\z2
-\frac{3}{8}\z3
-\frac{119}{48}\HPL_{0}
+\frac{3}{4}\z2\HPL_{0}
+\frac{7}{2}\z2\HPL_{-1}
-\frac{13}{12}\HPL_{0,0}
-\frac{33}{8}\HPL_{0,1}
-\frac{47}{24}\HPL_{-1,0}
\nonumber\\&&{}
-2\HPL_{-1,-1,0}
+\HPL_{0,-1,0}
-2\HPL_{0,1,0}
-4\HPL_{0,1,1}
+\frac{1}{4}\HPL_{0,0,1}
-\frac{5}{4}\HPL_{-1,0,0}
-\frac{9}{2}\HPL_{-1,0,1}
\Big)
-\frac{1}{24}(55+476\z2
\nonumber\\&&{}
+24\z3-816\z4)\delta(1-x)
\Big]
\nonumber\\&&{}
+\frac{32}{3}\cf^3\nf
\Big[
-41(1-x)
-(7+31x)\HPL_{0}
\nonumber\\&&{}
+24(1-x)\HPL_{1}
-4(2+x)\HPL_{0,0}
-12\pqqm\Big(
3\z2
+\frac{5}{2}\HPL_{0}
+2\HPL_{-1,0}
-\HPL_{0,0}
-2\HPL_{0,1}
\Big)
\nonumber\\&&{}
-2\pqqp\Big(
-\frac{37}{8}
+\frac{9}{2}\HPL_{0}
+\HPL_{0,0}
-6\HPL_{0,1}
-6\HPL_{1,0}
\Big)
+\frac{1}{8}(53+96\z2+80\z3)\delta(1-x)
\Big]
\nonumber\\&&{}
+\frac{64}{3}\cf^2\ca\nf\Big[
\frac{157}{4}(1-x)
+(1+x)\z2
+(9+22x)\HPL_{0}
-16(1-x)\HPL_{1}
\nonumber\\&&{}
+(3+2x)\HPL_{0,0}
-(1+x)\HPL_{0,1}
-\frac{1}{2}\pqqp\Big(
4\z2
-\frac{241}{12}
-\frac{29}{2}\HPL_{0}
-11\HPL_{0,0}
+12\HPL_{0,1}
+12\HPL_{1,0}
\Big)
\nonumber\\&&{}
+\pqqm(
24\z2+21\HPL_{0}
+16\HPL_{-1,0}
-7\HPL_{0,0}
-16\HPL_{0,1}
)
+\frac{1}{16}(235-104\z2-152\z3)\delta(1-x)
\Big]
\nonumber\\&&{}
+\frac{128}{3}\cf\ca^2\nf
\Big[
-\frac{359}{48}(1-x)
-3(1-x)\z2
\nonumber\\&&{}
-(1+x)\z3
-\frac{1}{8}(17+53x)\HPL_{0}
+6(1-x)\HPL_{1}
-(1+x)\z2\HPL_{0}
+3(1-x)\HPL_{0,1}
+3(1-x)\HPL_{1,0}
\nonumber\\&&{}
+6(1-x)\HPL_{1,1}
+(1+x)\HPL_{0,0,1}
+(1+x)\HPL_{0,1,0}
+2(1+x)\HPL_{0,1,1}
-\frac{1}{24}\pqqm
\Big(
89\z2
+\frac{159}{2}\HPL_{0}
\nonumber\\&&{}
+58\HPL_{-1,0}
-23\HPL_{0,0}
-60\HPL_{0,1}
\Big)
+\frac{1}{48}\pqqp
(
-385+58\z2-66\HPL_{0}
-82\HPL_{0,0}
)
-\frac{1}{192}(1789
\nonumber\\&&{}
-108\z2-216\z3)\delta(1-x)
\Big]
\nonumber\\&&{}
+\frac{128}{3}\,\frac{d_{44}^{\mathrm{RR}}}{\nr}\nf
\Big\{
44(1-x)
-72(1-x)\z2
-24(1+x)\z3
+12[1-9x-2\z2(1+x)]\HPL_{0}
\nonumber\\&&{}
+120(1-x)\HPL_{1}
-\pqqp(1-2\z2+2\HPL_{0,0})
-\pqqm(2\z2
+3\HPL_{0}
+4\HPL_{-1,0}
-2\HPL_{0,0})
\nonumber\\&&{}
+72(1-x)\HPL_{0,1}
+72(1-x)\HPL_{1,0}
+144(1-x)\HPL_{1,1}
+24(1+x)\HPL_{0,0,1}
+24(1+x)\HPL_{0,1,0}
\nonumber\\&&{}
+48(1+x)\HPL_{0,1,1}
-\frac{1}{4}(53-12\z2-24\z3)\delta(1-x)
\Big\}\,,
\nonumber\\
P_{\z3}^{(3)-}&=&
128\cf^4\Big\{
-\frac{424}{3}(1-x)
+\frac{1}{6}(511-123x)\z2
+16 x \z3
\nonumber\\&&{}
-\frac{1}{24}[1817-703x+24(9-25x)\z2]\HPL_{0}
-\frac{195}{2}(1-x)\HPL_{1}
-12(1+x)\z2\HPL_{-1}
-\frac{1}{6}(262+87x)\HPL_{0,0}
\nonumber\\&&{}
-\frac{1}{2}(91-41x)\HPL_{0,1}
-33(1-x)\HPL_{1,0}
+\frac{119}{3}(1+x)\HPL_{-1,0}
-60(1-x)\HPL_{1,1}
+(11-9x)\HPL_{0,0,0}
\nonumber\\&&{}
+5(3-5x)\HPL_{0,0,1}
+(1+x)\HPL_{0,1,0}
-3(1-x)\HPL_{1,0,0}
+6(1-x)\HPL_{0,-1,0}
+6(1+x)\HPL_{-1,0,0}
\nonumber\\&&{}
+2(1+x)\HPL_{0,1,1}
+12(1+x)\HPL_{-1,0,1}
-\pqqp\Big[
\frac{1}{8}(15-44\z2)\HPL_{0}
-\frac{15}{4}\HPL_{0,0}
+10\HPL_{0,0,0}
-6\HPL_{0,0,1}
\nonumber\\&&{}
+6\HPL_{1,0,0}
-20\HPL_{0,-1,0}-30\z3\Big]
-\pqqm\Big[
\frac{3}{2}(\z2-41\z3)
-\frac{1}{6}(8+411\z2)\HPL_{0}
+42\z2\HPL_{-1}
-\frac{21}{4}\HPL_{0,0}
\nonumber\\&&{}
-\frac{3}{2}\HPL_{0,1}
-27\HPL_{-1,0,0}
-42\HPL_{-1,0,1}
-3\HPL_{0,-1,0}+26\HPL_{0,0,0}
+57\HPL_{0,0,1}+
33\HPL_{0,1,0}+60\HPL_{0,1,1}
\Big]
\nonumber\\&&{}
+\frac{1}{24}\delta(1-x)(
315-58\z2
-324\z3+108\z4)
\Big\}
\nonumber\\&&{}
+64\cf^3\ca\Big\{
\frac{8003}{12}(1-x)
-\frac{1}{6}(2191 - 477 x)\z2
\nonumber\\&&{}
-(13 + 53 x) \z3
+\frac{1}{6}[2205 - 537 x+4(39 - 120 x) \z2]\HPL_{0}
+\frac{891}{2}(1-x)\HPL_{1}
+60(1+x)\z2\HPL_{-1}
\nonumber\\&&{}
+\frac{1}{6}(1130 + 383 x)\HPL_{0,0}
+\frac{1}{2}(365 - 159 x)\HPL_{0,1}
+131(1-x)\HPL_{1,0}
-\frac{548}{3}(1+x)\HPL_{-1,0}
+244(1-x)\HPL_{1,1}
\nonumber\\&&{}
-\frac{1}{2}(67 - 45 x)\HPL_{0,0,0}
-2(22 - 40 x)\HPL_{0,0,1}
-12(1+x)\HPL_{0,1,0}
+15(1-x)\HPL_{1,0,0}
-24(1+x)\HPL_{0,1,1}
\nonumber\\&&{}
-18(1-x)\HPL_{0,-1,0}
-30(1+x)\HPL_{-1,0,0}
-60(1+x)\HPL_{-1,0,1}
-\pqqm\Big[
\frac{1}{2}(71\z2+445\z3)
\nonumber\\&&{}
+\frac{1}{6}(206+1581\z2)\HPL_{0}
-134\z2\HPL_{-1}
+\frac{27}{4}\HPL_{0,0}
-\frac{71}{2}\HPL_{0,1}
-90\HPL_{0,0,0}-207\HPL_{0,0,1}
-131\HPL_{0,1,0}
\nonumber\\&&{}
+9\HPL_{0,-1,0}
+85\HPL_{-1,0,0}
+134\HPL_{-1,0,1}
-244\HPL_{0,1,1}
\Big]
-\pqqp\Big[148\z3\
+\frac{3}{8}(76\z2-15)\HPL_{0}
+\frac{113}{12}\HPL_{0,0}
\nonumber\\&&{}
-48\HPL_{0,0,0}
+104\HPL_{0,-1,0}
+28\HPL_{0,0,1}-20\HPL_{1,0,0}
\Big]
-\frac{1}{24}\delta(1-x)(739+554\z2-1760\z3+432\z4)
\Big\}
\nonumber\\&&{}
+128\cf^2\ca^2\Big\{
-\frac{11849}{48}(1-x)
+\frac{1}{6}(803 - 178 x)\z2
+\frac{1}{4} (17 + 59 x)\z3
-\frac{1}{48}[6751 - 685 x+24 (20 
\nonumber\\&&{}
- 43 x) \z2]\HPL_{0}
-\frac{1913}{12}(1-x)\HPL_{1}
-24(1+x)\z2\HPL_{-1}
-\frac{1}{24}(1703 + 513 x)\HPL_{0,0}
-\frac{89}{3}(2 - x)\HPL_{0,1}
\nonumber\\&&{}
-\frac{89}{2}(1-x)\HPL_{1,0}
+\frac{149}{2}(1+x)\HPL_{-1,0}
-86(1-x)\HPL_{1,1}
+3(3-1x)\HPL_{0,0,0}
+\frac{1}{2}(29 - 43 x)\HPL_{0,0,1}
\nonumber\\&&{}
+\frac{17}{2}(1+x)\HPL_{0,1,0}
-6(1-x)\HPL_{1,0,0}
+17(1+x)\HPL_{0,1,1}
+\frac{9}{2}(1-x)\HPL_{0,-1,0}
+12(1+x)\HPL_{-1,0,0}
\nonumber\\&&{}
+24(1+x)\HPL_{-1,0,1}
-\pqqm\Big[
-\frac{1}{4}(82\z2+279\z3)
-\frac{1}{4}(85+343\z2)\HPL_{0}
+35\z2\HPL_{-1}
-\frac{35}{24}\HPL_{0,0}
\nonumber\\&&{}
+\frac{275}{12}\HPL_{0,1}
+\frac{29}{6}\HPL_{-1,0}
+25\HPL_{0,0,0}
+\frac{125}{2}\HPL_{0,0,1}
+\frac{89}{2}\HPL_{0,1,0}
-\frac{7}{2}\HPL_{0,-1,0}
-\frac{43}{2}\HPL_{-1,0,0}
+86\HPL_{0,1,1}
\nonumber\\&&{}
-33\HPL_{-1,0,1}
+4\HPL_{-1,-1,0}
\Big]
-\pqqp\Big[
\frac{1}{12}(29\z2-714\z3)
+\frac{1}{16}(15-212\z2)\HPL_{0}
-2\z2\HPL_{1}
+\frac{1}{24}\HPL_{0,0}
\nonumber\\&&{}
+19\HPL_{0,0,0}
-10\HPL_{0,0,1}
+5\HPL_{1,0,0}
-43\HPL_{0,-1,0}
\Big]
-\frac{1}{192}\delta(1-x)(
411 - 3720 \z2 + 5672 \z3 - 2040 \z4)\Big\}
\nonumber\\&&{}
+\frac{32}{3}\cf\ca^3\Big\{
\frac{2000}{3}(1-x)
-\frac{1}{6}(2087-339x)\z2
+\frac{1}{2}(3-49x)\z3
+\frac{1}{12}[ 4897 + 751 x + (546 -474 x)\z2]\HPL_{0}
\nonumber\\&&{}
+\frac{757}{2} (1 - x) \HPL_{1}
+78(1+x)\z2\HPL_{-1}
+\frac{1}{6}(1265 + 342 x)\HPL_{0,0}
+\frac{1}{2}(207 -113 x)\HPL_{0,1}
+80(1-x)\HPL_{1,0}
\nonumber\\&&{}
+160(1-x)\HPL_{1,1}
-\frac{733}{3}(1+x)\HPL_{-1,0}
-18\HPL_{0,0,0}
-\frac{1}{2}(111-79x)\HPL_{0,0,1}
-47(1+x)\HPL_{0,1,0}
\nonumber\\&&{}
+\frac{39}{2}(1-x)\HPL_{1,0,0}
-94(1+x)\HPL_{0,1,1}
-10(1-x)\HPL_{0,-1,0}
-39(1+x)\HPL_{-1,0,0}
-78(1+x)\HPL_{-1,0,1}
\nonumber\\&&{}
-\pqqm\Big[
\frac{1}{12}(889\z2+2106\z3)
+\frac{5}{12}(163+540\z2)\HPL_{0}
-58\z2\HPL_{-1}
+\frac{31}{6}\HPL_{0,0}
-\frac{221}{6}\HPL_{-1,0}
-\frac{185}{2}\HPL_{0,1}
\nonumber\\&&{}
-32\HPL_{-1,-1,0}
+37\HPL_{-1,0,0}
+42\HPL_{-1,0,1}
+16\HPL_{0,-1,0}
-54\HPL_{0,0,0}
-149\HPL_{0,0,1}
-128\HPL_{0,1,0}
-256\HPL_{0,1,1}
\Big]
\nonumber\\&&{}
-\pqqp\Big[
\frac{1}{36}(-419-663\z2+6390\z3)
+47\z2\HPL_{0}
+16\z2\HPL_{1}
-\frac{59}{6}\HPL_{0,0}
-54\HPL_{0,0,0}
+29\HPL_{0,0,1}
\nonumber\\&&{}
-\HPL_{1,0,0}
+136\HPL_{0,-1,0}
\Big]
+\frac{1}{24}\delta(1-x)(1351
-1630\z2+2268\z3-1452\z4)
\Big\}
\nonumber\\&&{}
+128\frac{d_{44}^{\mathrm{RA}}}{\nr}
\Big\{
-\frac{28}{3}(1-x)
+\frac{1}{3}(80-159x)\z2
+(9+7x)\z3
\nonumber\\&&{}
-\frac{1}{6}[67-53x-12(1+5x)\z2]\HPL_{0}
-20(1-x)\HPL_{1}
-12(1+x)\z2\HPL_{-1}
-\frac{1}{3}(5-9x)\HPL_{0,0}
\nonumber\\&&{}
-(3-53x)\HPL_{0,1}
-28(1-x)\HPL_{1,0}
-56(1-x)\HPL_{1,1}
+\frac{71}{3}(1+x)\HPL_{-1,0}
-10x\HPL_{0,0,1}
+(1+x)\HPL_{0,1,0}
\nonumber\\&&{}
-3(1-x)\HPL_{1,0,0}
+2(1-x)\HPL_{0,-1,0}
+6(1+x)\HPL_{-1,0,0}
+2(1+x)\HPL_{0,1,1}
+12(1+x)\HPL_{-1,0,1}
\nonumber\\&&{}
-\pqqm\Big[
\frac{1}{6}(121-36\z2)\HPL_{0}
-28\z2\HPL_{-1}
+\frac{8}{3}\HPL_{0,0}
+21\HPL_{0,1}
+\frac{47}{3}\HPL_{-1,0}
-2\HPL_{0,0,1}
+16\HPL_{0,1,0}
\nonumber\\&&{}
+32\HPL_{0,1,1}
-8\HPL_{0,-1,0}
+10\HPL_{-1,0,0}
+36\HPL_{-1,0,1}
+16\HPL_{-1,-1,0}
-\frac{79}{6}\z2+3\z3
\Big]
\nonumber\\&&{}
-\pqqp\Big(
-\frac{1}{6}+\frac{47}{6}\z2-\z3
-10\z2\HPL_{0}-8\z2\HPL_{1}-8\HPL_{0,-1,0}+
2\HPL_{0,0,1}-10\HPL_{1,0,0}
+\frac{26}{3}\HPL_{0,0}
\Big)
\nonumber\\&&{}
-\frac{1}{24}\delta(1-x)(55+484\z2+24\z3-816\z4)
\Big\}
\nonumber\\&&{}
+128\cf^3\nf\Big[
-\frac{35}{12}(1-x)
-\frac{19}{12}(1+x)\HPL_{0}
-2(1-x)\HPL_{1}
-\frac{1}{3}(1+2x)\HPL_{0,0}
\nonumber\\&&{}
+\pqqm\Big(
3\z2
+\frac{5}{2}\HPL_{0}
-\HPL_{0,0}
-2\HPL_{0,1}
+2\HPL_{-1,0}
\Big)
+\pqqp\Big(
\frac{37}{48}
-\frac{3}{4}\HPL_{0}
-\frac{1}{6}\HPL_{0,0}
\nonumber\\&&{}
+\HPL_{0,1}
+\HPL_{1,0}
\Big)
+\frac{1}{96}\delta(1-x)(53+96\z2+80\z3)
\Big]
\nonumber\\&&{}
+\frac{64}{3}\cf^2\ca\nf\Big[
\frac{119}{4}(1-x)
\nonumber\\&&{}
+(1+x)\z2
+(14+13x)\HPL_{0}
+16(1-x)\HPL_{1}
+(2+3x)\HPL_{0,0}
-(1+x)\HPL_{0,1}
-\pqqm(
24\z2
\nonumber\\&&{}
+21\HPL_{0}
+16\HPL_{-1,0}
-7\HPL_{0,0}
-16\HPL_{0,1}
)
+\pqqp\Big(
\frac{241}{24}
-2\z2
+\frac{29}{4}\HPL_{0}
+\frac{11}{2}\HPL_{0,0}
-6\HPL_{0,1}
-6\HPL_{1,0}
\Big)
\nonumber\\&&{}
+\frac{1}{16}\delta(1-x)(235-104\z2-152\z3)
\Big]
\nonumber\\&&{}
+\frac{128}{3}
\cf\ca^2\nf\Big\{
-\frac{79}{16}(1-x)
-3(1-x)\z2
-(1+x)\z3
\nonumber\\&&{}
-\frac{1}{24}[55+115x+24(1+x)\z2]\HPL_{0}
+(1-x)\HPL_{1}
+3(1-x)\HPL_{0,1}
+3(1-x)\HPL_{1,0}
\nonumber\\&&{}
+6(1-x)\HPL_{1,1}
+(1+x)\HPL_{0,0,1}
+(1+x)\HPL_{0,1,0}
+2(1+x)\HPL_{0,1,1}
-\pqqm\Big(
-\frac{89}{24}\z2-\frac{53}{16}\HPL_{0}
\nonumber\\&&{}
+\frac{23}{24}\HPL_{0,0}
+\frac{5}{2}\HPL_{0,1}
-\frac{29}{12}\HPL_{-1,0}\Big)
-\pqqp\Big(
\frac{1}{48}(385-58\z2)+\frac{11}{8}\HPL_{0}+
\frac{41}{24}\HPL_{0,0}
\Big)
\nonumber\\&&{}
-\frac{1}{192}\delta(1-x)(1789-108\z2-216\z3)
\Big\}
\nonumber\\&&{}
+1024 \frac{d_{44}^{\mathrm{RR}}}{\nr}n_f
\Big\{\frac{3}{4}(1-x)-3(1-x)\z2-(1+x)\z3
-\frac{1}{6}[1+31x+6(1+x)\z2]\HPL_{0}
\nonumber\\&&{}
+5(1-x)\HPL_{1}
+3(1-x)\HPL_{0,1}
+3(1-x)\HPL_{1,0}
+6(1-x)\HPL_{1,1}
+(1+x)\HPL_{0,0,1}
\nonumber\\&&{}
+(1+x)\HPL_{0,1,0}
+2(1+x)\HPL_{0,1,1}
+\frac{1}{24}\pqqm(
2\z2
+3\HPL_{0}
-2\HPL_{0,0}
+4\HPL_{-1,0}
)
-\frac{1}{24}\pqqp(
1-2\z2
\nonumber\\&&{}
+2\HPL_{0,0}
)
+\frac{1}{96}\delta(1-x)(-53+12\z2+24\z3)\Big\}\,,
\label{eq:padN}
\end{eqnarray}
\normalsize
where the argument $x$ has been dropped in the HPLs.

In the limit $x\to0$, we have
\footnotesize
\begin{eqnarray}
P_{\z3}^{(3)+}&=&
\frac{16}{3} ( 28 \cf^4 - 39  \cf^3\ca + 12  \cf^2\ca^2)\ln^3x
+\frac{8}{3} (
192 \cf^4 
- 226 \cf^3\ca 
- 9  \cf^2\ca^2
+33 \cf\ca^3 
\nonumber\\&&{}
+ 4 \cf^3 \nf
+ 6 \cf^2\ca  \nf
-  6  \cf\ca^2 \nf
)\ln^2x
+\frac{8}{3} \Big[
12 \cf^4 (71 - 152 \z2)
+ \cf^3\ca  (643 + 3720 \z2) 
\nonumber\\&&{}
- 12  \cf^2\ca^2 (193 + 204 \z2) 
+ 2   \cf\ca^3 (423 + 277 \z2)
+ 48 (15 + 14 \z2) \frac{d_{44}^{\mathrm{RA}}}{\nr}
- 184 \cf^3 \nf
\nonumber\\&&{}
 +298 \cf^2\ca  \nf
-  \cf\ca^2 \nf (109 + 16 \z2)
+ 48 \frac{d_{44}^{\mathrm{RR}}}{\nr}\nf(3 - 8 \z2) 
\Big]\ln x
\nonumber\\&&{}
+\frac{8}{27} \Big[
72 \cf^4 (226 - 150 \z2 - 93 \z3) 
- 18 \cf^3\ca  (295 - 1404 \z2 - 258 \z3) 
\nonumber\\&&{}
- 9  \cf^2\ca^2 (1947 + 1804 \z2 - 216 \z3) 
+  \cf\ca^3 (6533 + 3636 \z2 - 954 \z3) 
\nonumber\\&&{}
+ 216 \frac{d_{44}^{\mathrm{RA}}}{\nr}(93 + 140 \z2 + 22 \z3)  
-9 \cf^3 \nf (127 + 144 \z2) 
+ 3 \cf^2\ca  \nf (1183 + 552 \z2) 
\nonumber\\&&{}
- 72 \cf\ca^2  \nf (31 + 11 \z2 + 2 \z3) 
+ 144 \frac{d_{44}^{\mathrm{RR}}}{\nr}\nf(43 - 72 \z2 - 24 \z3) 
\Big]\,,
\nonumber\\
P_{\z3}^{(3)-}&=&
\frac{16}{3} (- 100 \cf^4 + 209 \cf^3\ca  - 140  \cf^2\ca^2 +30  \cf\ca^3)\ln^3x
+\frac{8}{9} \Big(
- 2496 \cf^4 
+ 6198 \cf^3\ca  
\nonumber\\&&{}
- 5007 \cf^2\ca^2 
+1293 \cf\ca^3 
- 936 \frac{d_{44}^{\mathrm{RA}}}{\nr} 
- 108 \cf^3 \nf 
+ 174 \cf^2\ca  \nf
- 64  \cf\ca^2 \nf
\nonumber\\&&{}
- 96 \frac{d_{44}^{\mathrm{RR}}}{\nr}\nf\Big)\ln^2x 
+\frac{8}{9} \Big[
  - 180 \cf^4 (61 - 52 \z2)
+ 3 \cf^3\ca  (8131 - 6384\z2)
\nonumber\\&&{}
  - 48 \ca^2 \cf^2 (361 - 267 \z2) 
+ 2 \ca^3 \cf (2041 - 1359 \z2)
- 96 \frac{d_{44}^{\mathrm{RA}}}{\nr}(47 - 27 \z2)  
  + 24 \cf^3 \nf
\nonumber\\&&{}
+  6 \cf^2\ca  \nf
  - \cf \ca^2 \nf (17 + 48 \z2) 
- 48 \frac{d_{44}^{\mathrm{RR}}}{\nr}\nf(1 + 24 \z2) 
\Big]\ln x
\nonumber\\&&{}
+\frac{8}{27} \Big[
- 72 \cf^4 (848 - 502 \z2 - 549 \z3) 
+ 18 \cf^3\ca  (8003 - 4808 \z2 - 4602 \z3) 
\nonumber\\&&{}
- 9 \cf^2\ca^2  (11849 - 7292 \z2 - 6408 \z3) 
+ \cf\ca^3  (24419 - 14526 \z2 - 12654 \z3) 
\nonumber\\&&{}
- 72 \frac{d_{44}^{\mathrm{RA}}}{\nr} (55 - 192 \z2 - 42 \z3) 
- 9 \cf^3 \nf (103 - 144 \z2) 
  +15 \cf^2\ca  \nf (191 - 120 \z2) 
\nonumber\\&&{}
- 6 \cf\ca^2  \nf (311 - 46 \z2 + 24 \z3) 
+ 144 \frac{d_{44}^{\mathrm{RR}}}{\nr}\nf(17 - 68 \z2 - 24 \z3) \Big]\,.
\end{eqnarray}
\normalsize

In the limit $x\to1$, we have
\footnotesize
\begin{eqnarray}
  P_{\z3}^{(3)+}&=&512\zeta_2\Big\{
  \Big[\frac{\ln(1-x)}{1 - x}\Big]_+-\ln(1-x)\Big\}
  \Big(-\cf^2\ca^2+\frac{2}{3}\cf\ca^3-4\frac{d_{44}^{\mathrm{RA}}}{\nr}\Big)
  \nonumber\\&&{}
  +\frac{16}{3}\Big[\frac{1}{1 - x}\Big]_+
  \Big[
    -116\cf^2\ca^2\z2
    +\frac{1}{9}\cf\ca^3(419 + 663 \z2 - 54 \z3)
    +8\frac{d_{44}^{\mathrm{RA}}}{\nr}(1-47\z2-18\z3)
    \nonumber\\&&{}  
    +37\cf^3\nf
    +\frac{1}{3}\cf^2\ca\nf(241 - 48 \z2)
    -\frac{1}{3}\cf\ca^2\nf(385 - 58 \z2)
    -16\frac{d_{44}^{\mathrm{RR}}}{\nr}\nf(1 - 2 \z2)\Big] 
  \nonumber\\&&{}
  +\frac{2}{3}\delta(1 - x)\Big[
    8\cf^4 (315 - 50 \z2 - 324 \z3 + 108 \z4)
    -4\cf^3\ca (739 + 586 \z2- 1760 \z3 + 432 \z4)
 \nonumber\\&&{}   
- \cf^2\ca^2(411 - 3816 \z2 + 5672 \z3 - 2040 \z4)
+\frac{2}{9}  \cf\ca^3(4053 - 5010 \z2 + 6804 \z3- 4356 \z4)
  \nonumber\\&&{}
-8\frac{d_{44}^{\mathrm{RA}}}{\nr}(55+476 \z2+24 \z3- 816 \z4)
+2 \cf^3  \nf(53+96 \z2 - 80 \z3)
+\frac{2}{3} \cf^2\ca  \nf(705 - 312 \z2 - 456 \z3)
  \nonumber\\&&{}
- \frac{1}{3} \cf \ca^2\nf(1789- 108 \z2- 216 \z3)
- 16\frac{d_{44}^{\mathrm{RR}}}{\nr}\nf(53 - 12 \z2 - 24 \z3) \Big]
\nonumber\\&&{}
+\frac{1856}{3}\z2\cf^2\ca^2 
-\frac{16}{27}\cf\ca^3(419+663\z2-54\z3)
-\frac{128}{3}\,\frac{d_{44}^{\mathrm{RA}}}{\nr}(1-47\z2-18\z3)
  \nonumber\\&&{}
-\frac{592}{3}\cf^3\nf
-\frac{16}{9}\cf^2\ca\nf(241-48\z2)
+\frac{16}{9}\cf\ca^2\nf(385-58\z2)
+\frac{256}{3}\,\frac{d_{44}^{\mathrm{RR}}}{\nr}\nf(1-2\z2)\,,
\nonumber\\
P_{\z3}^{(3)-}&=&P_{\z3}^{(3)+}+32\z2
\Big(- \frac{4}{3} \cf^4 
+ \frac{8}{3} \cf^3\ca 
- 2  \cf^2\ca^2
+\frac{5}{9}  \cf\ca^3
- \frac{4}{3}\,\frac{d_{44}^{\mathrm{RA}}}{\nr}\Big)\delta(1-x)\,.
\end{eqnarray}

\end{document}